%% file: OAML_HeavyTailedNoise_JSTSP_R1Rvs_v1.tex
\documentclass[twocolumn, 10pt]{IEEEtran}
\usepackage{graphicx}
\usepackage{amssymb}
\usepackage{amsmath}
\usepackage{mathtools}
\usepackage{dsfont}
\usepackage{cite}
\usepackage{stfloats}
\usepackage{subfigure}
\usepackage{psfrag}
\usepackage[mathscr]{euscript}
\usepackage{algorithm, algorithmic}
\usepackage{acronym}  
\usepackage{booktabs}
\usepackage{bbm}

\usepackage{float}
\usepackage{balance}
\input{SupportDocuments/aconym}
\input{SupportDocuments/defmetric}

\newcommand{\paperTitle}{ Revisiting Analog Over-the-Air Machine Learning: The Blessing and Curse of Interference }

\begin{document}

{
\title{\paperTitle}

\author{
    
    Howard H. Yang, \textit{Member, IEEE},
    Zihan~Chen, \textit{Student Member, IEEE},\\
    Tony Q. S. Quek, \textit{Fellow, IEEE},
    and H. Vincent Poor, \textit{Fellow, IEEE}



\thanks{H. H. Yang is with the Zhejiang University/University of Illinois at Urbana-Champaign Institute, Zhejiang University, Haining 314400, China, the College of Information Science and Electronic Engineering, Zhejiang University, Hangzhou 310007, China, and the Department of Electrical and Computer Engineering, University of Illinois at Urbana-Champaign, Champaign, IL 61820, USA (email: haoyang@intl.zju.edu.cn).}

\thanks{Z. Chen and T.~Q.~S.~Quek are with the Information Systems Technology and Design Pillar, Singapore University of Technology and Design, Singapore (e-mail: zihan\_chen@mymail.sutd.edu.sg, tonyquek@sutd.edu.sg).}
%
%

\thanks{H.~V.~Poor is with the Department of Electrical and Computer Engineering, Princeton University, Princeton, NJ 08544 USA (e-mail: poor@princeton.edu).}
}
\maketitle
\acresetall
\thispagestyle{empty}
\begin{abstract}
We study a distributed machine learning problem carried out by an edge server and multiple agents in a wireless network. The objective is to minimize a global function that is a sum of the agents' local loss functions.
And the optimization is conducted by analog over-the-air model training.
Specifically, each agent modulates its local gradient onto a set of waveforms and transmits to the edge server simultaneously.
From the received analog signal the edge server extracts a noisy aggregated gradient which is distorted by the channel fading and interference, and uses it to update the global model and feedbacks to all the agents for another round of local computing.
Since the electromagnetic interference generally exhibits a heavy-tailed intrinsic, we use the $\alpha$-stable distribution to model its statistic.
In consequence, the global gradient has an infinite variance that hinders the use of conventional techniques for convergence analysis that rely on second-order moments' existence. 
To circumvent this challenge, we take a new route to establish the analysis of convergence rate, as well as generalization error, of the algorithm. 
We also show that the training algorithm can be run in tandem with the momentum scheme to accelerate the convergence. 
Our analyses reveal a two-sided effect of the interference on the overall training procedure. 
On the negative side, heavy tail noise slows down the convergence rate of the model training: the heavier the tail in the distribution of interference, the slower the algorithm converges.
On the positive side, heavy tail noise has the potential to increase the generalization power of the trained model: the heavier the tail, the better the model generalizes. 
This perhaps counterintuitive conclusion implies that the prevailing thinking on interference -- that it is only detrimental to the edge learning system -- is outdated and 
we shall seek new techniques that exploit, rather than simply mitigate, the interference for better machine learning in wireless networks.
\end{abstract}
\acresetall

\section{Introduction}\label{sec:intro}
We consider a distributed machine learning problem conducted in a mobile edge network. Particularly, a group of $N$ agents communicate over the spectrum to an edge server, whereas each agent has a local objective function $f_n: \mathbb{R}^d \rightarrow \mathbb{R}$, and the goal is to minimize the global loss function:
\begin{align}
    f(\boldsymbol{w}) = \frac{1}{N} \sum_{n=1}^N f_n(\boldsymbol{w}).
\end{align}
Due to privacy concerns, the agents do not share their data, and the minimization can only be carried out in a decentralized manner. 
To that end, we adopt analog over-the-air model training \cite{SerCoh:20TSP}, which is mainly based on the gradient descent (GD) method, to accomplish this task. Specifically, each agent calculates its local gradient and modulates it onto $d$ orthonormal waveforms, one for each element of the gradient vector. Then, the agents send out their analog signals simultaneously.
The edge server receives the superposition of the analog signals, which represents a noisy global gradient distorted by the channel fading and interference. 
Based on this noisy gradient, the edge server updates the global parameter and feeds it back to all the agents for another round of local computation.
The procedure repeats until the model converges. 

We inspect this algorithm in a more pragmatic and complicated setting where interference follows an $\alpha$-stable distribution \cite{ClaPedRod:20}. 
In that context, the variance of the aggregated gradient is infinite, and the effects of such a phenomenon on the training procedure, or the algorithm can even converge or not, remain unknown.
The central thrust of the present article is to fill this research gap.

\subsection{Main Contributions}
This paper builds upon the model in \cite{SerCoh:20TSP} but differs from it by considering the heavy-tailed nature of the electromagnetic interference \cite{Mid:77}. 
Specifically, we adopt the symmetric $\alpha$-stable distribution -- a widely used model in wireless networks \cite{Hae:09,WinPin:09} -- to model the statistics of interference.
The parameter $\alpha$ is commonly known as the \textit{tail index} where smaller the $\alpha$ means heavier the tail in the distribution. 
Under such a setting, the aggregated global gradient admits diverging variance, and the conventional approaches that heavily rely on the existence of second moments for convergence analysis fail to function. 
In that respect, we take a new route toward the convergence analysis and verify that even the intermediate gradients are severely distorted by channel fading and interference, GD-based methods can ultimately reach the optimal solution. 

Our main contributions are summarized as follows:
\begin{itemize}
    \item We derive analytical expression to characterize the convergence rate of the analog over-the-air GD algorithm, which encompasses key sytem parameters such as the number of agents, channel fading, and interference. Particularly, the convergence rate is in the order of $\mathcal{O}(1/k^{\alpha-1})$, where $k$ stands for the communication round. This result also implies that heavier tailed interference leads to slower convergence of the algorithm.  
    \item We show that analog over-the-air GD can be run in conjunction with momentum. We also derive the convergence rate by taking into account the momentum weight. Our result reveals that the momentum based model training also converges in the order of $\mathcal{O}(1/k^{\alpha-1})$, while the momentum weight affects the multiplication factor.
    \item We analytically characterize the generalization error of analog over-the-air GD by resorting to a continuous time proxy of the update trajectory. The analysis shows that heavy tail can potentially improve the algorithm's generalization capability. More precisely, with a certain probability, the generalization error decreases along with the tail index.
    \item We conduct extensive simulations on the MNIST and CIFAR-10 data set to examine the algorithm under different system parameters. The experiments demonstrate that an increase in the number of agents, learning rate, or tail index of the interference leads to a faster convergence rate. It also shows that occasionally a smaller tail index results in better prediction accuracy of the trained model, which confirms that heavy tail has the potential to improve the generalization capability.
\end{itemize}

\subsection{Prior Art}
Distributed optimizations in wireless networks have garnered considerable attention in recent years, especially with the rise of federated learning \cite{MaMMooRam:17AISTATS,LiSahTal:20SPM,ParSamBen:19}. 
The typical system is generally constituted by an edge server and a number of agents, where the goal is to collaboratively optimize an objective function via the orchestration amongst the network elements.
Particularly, each agent conducts on-device training based on its local dataset, and uploads the intermediate result, e.g., the gradient, to the server for model improvement. Then, they download the new model for another round of local computing. This procedure repeats multiple rounds until the training converges. 
Upon each global iteration, the transmissions of model parameters need to go over the spectrum, which is resource-limited and unreliable. 
Recognizing the conventional schemes that hinged on the separated communication-and-computation principle can encounter difficulty in accomondating massive access and stringent latency requirements, a recent line of studies \cite{ZhuXuHua:21MAG} proposed utilizing the over-the-air computing to enable efficient model aggregation and hence achieve faster machine learning over many devices.  

The essence of over-the-air computing is to exploit the waveform superposition property of multi access channel, where agents modulate the gradient on the waveform and use the air as an auto aggregator. 
In the presence of channel fading, it is suggested to invert the channel via power control at the end-user devices where the nodes that encounter deep fades suspend their transmissions \cite{ZhuWanHua:19TWC,YanJiaShi:20}. 
And the server shall adopt better scheduling methods in each communication round to rev up the model training process. 
To reduce communication overheads, the devices can compress the gradient vectors by sending out a sparse  \cite{AmiGun:20TSP}, or even a one-bit quantized \cite{ZhuDuGun:21TWC}, version, followed by QAM modulation. 
At the edge server side, it can expand the antenna array to further mitigate the effects of channel fading, where the fading vanishes as the spatial dimension approaches infinity \cite{AmiDumGun:21TWC}. 
Furthermore, \cite{SerShlCoh:20} devise a precoding scheme that gradually amplifies the model updates as the training progresses to handle the performance degradation incurred by the additive noise. 
With the help of feedbacks, \cite{GuoLiuLau:20} optimizes the transceiver parameters by jointly accounting for the data and channel states to cope with the nonstationality of the gradient updates. 
Inspired by the fact that machine learning algorithms need not to operate under impeccably precise parameters, the authors of \cite{SerCoh:20TSP} suggest the agents directly transmit the analog gradient signals without any power control or beamforming to invert the channel whilst the server updates the global model based on the noisy gradient. 
They also show that the convergence is guaranteed.
This approach substantially simplifies the system design while achieves virtually zero access latency \cite{CaiLau:17}.
What is more appealing, the data privacy is in fact enhanced by implicitly harnessing the randomness of wireless medium and the training procedure can be accelerated by adopting an analog ADMM-type algorithm \cite{ElgParIss:21}.
Despite the wealth of work in this area, a significant restriction in almost all the previous results lies at the presumption that the interference follows a normal distribution. 
While convenient, this assumption hardly holds in practice as the constructive property of the electromagnetic waves often results in heavy tails in the distribution of interference \cite{Mid:77,Hae:09,WinPin:09}.
In consequence, there is a non-negligible chance that the magnitude of interference sheers to a humungous value in some communication rounds which wreaks havoc on the global model. 
Understanding the impact of such a phenomenon on the performance of the learning algorithm is the focus of this work.

In this paper, we use bold lower case letters to denote column vectors. For any vector $\boldsymbol{w} \in \mathbb{R}^d$, we use $\Vert \boldsymbol{w} \Vert$ and $\boldsymbol{w}^{\mathsf{T}}$ to denote the $L$-2 norm and the transpose of a column vector, respectively. The main notations used throughout the paper are summarized in Table~I.

The remainder of this paper is organized as follows. We introduce the system model in Section II.
In Section III, we derive the convergence rate of analog over-the-air GD.
We also present the convergence rate of analog over-the-air GD with momentum.
In Section IV, we analyze the generalization error of the analog over-the-air GD algorithm.
Then, we show the simulation results in Section V to validate the analyses and obtain design insights.
We conclude the paper in Section VI.

 \begin{table}
 \caption{ Notation Summary } \label{table:notation}
 \begin{center}
 \renewcommand{\arraystretch}{1.3}
 \begin{tabular}{c  p{ 5.5cm } }
 \hline
  {\bf Notation} & {\hspace{2.5cm}}{\bf Definition}
 \\
 \hline
 $N$; $\mathbf{s}(t)$ & Number of clients in the network; a set of orthonormal waveforms \\
 $f( \boldsymbol{w} )$; $\nabla f( \boldsymbol{w} )$ & Global loss function; and its gradient \\
 $f_n( \boldsymbol{w} )$; $\nabla f_n( \boldsymbol{w} )$ & Local loss function of client $n$; and its gradient \\
 $x_{n}(t)$; $y(t)$ & Analog signal sent out by client $n$; analog signal received by the server \\
 $P_{n}$; $h_{n,k}$ & Transmit power of client $n$; channel fading experience by client $n$ \\
 $\boldsymbol{g}_k$; $\boldsymbol{\xi}_k$ & Noisy gradient received at the server; electromagnetic interference that follows $\alpha$-stable distribution \\
 $\alpha$; $\beta$ & Tail index of the heavy-tailed interference; controlling factor, a.k.a., momentum weight, of the momentum algorithm \\
 $\eta_k$ & Learning rate of the GD-based training algorithm \\
 $\boldsymbol{w}^{\langle \alpha \rangle}$; $\Vert \boldsymbol{w} \Vert_\alpha$  & Signed power $\alpha$ of a vector $\boldsymbol{w}$; $\alpha$-norm of a vector $\boldsymbol{w}$ \\
$\mathcal{R}( \boldsymbol{w} )$; $\hat{ \mathcal{R} } ( \boldsymbol{w}, \mathcal{D} )$  & Population risk of the machine learning task; empirical risk of the machine learning task constituted from dataset $\mathcal{D}$ \\ 
$\mathcal{I}( \boldsymbol{w}_K )$ & Generalization error of the algorithm \\
 \hline
 \end{tabular}
 \end{center}\vspace{-0.63cm}
 \end{table}%


\section{System Model}

\subsection{Setting}

Let us consider an edge learning system consisting of one server and $N$ agents. 
Each agent $n$ holds a local dataset $\mathcal{D}_n = \{ ( \boldsymbol{x}_i \in \mathbb{R}^d, y_i \in \mathbb{R} ) \}_{ i = 1 }^{m_n}$ with size $\vert \mathcal{D}_n \vert = m_n$, and we assume the local datasets are statistically independent across the clients.
The goal of all the entities in this system is to jointly learn a statistical model constituted from all the data samples of the clients. 
More precisely, they need to find a vector $\boldsymbol{w} \in \mathbb{R}^d$ that minimizes a global loss given as follows:
\begin{align} \label{equ:ObjFunc}
f( \boldsymbol{w} ) = \frac{1}{N} \sum_{ n=1 }^N f_n( \boldsymbol{w} )
\end{align}
where $f_n ( \boldsymbol{w} )$ is the local empirical risk of agent $n$, given by
\begin{align}
f_n ( \boldsymbol{w} ) = \frac{1}{ m_n } \sum_{ i=1 }^{ m_n } \ell( \boldsymbol{w}; \boldsymbol{x}_i, y_i ).
\end{align}
The solution is commonly known as the empirical risk minimizer, denoted by
\begin{align}
\boldsymbol{w}^* = \arg\min f(\boldsymbol{w}).
\end{align}

Due to privacy concerns, the agents are unwilling to share their local dataset and hence the minimization of \eqref{equ:ObjFunc} needs to be conducted by means of distributed learning. 
Particularly, the agents minimize their local loss and upload the intermediate gradients to the server, with which the server conducts a global aggregation to improve the model and feeds it back to the agents for another round of local training. 
Such interactions between the server and agents repeat until the model converges.
During this process, we consider the communications amongst the server and the agents are taken place over the spectrum, which is by nature resource-limited and unreliable.
In light of its efficacy in spectral utilization, we adopt the analog over-the-air computing \cite{NazGas:07IT} for the training of the statistical model, which is detailed in the sequel. 

\begin{figure}[t!]
  \centering{}

    {\includegraphics[width=0.98\columnwidth]{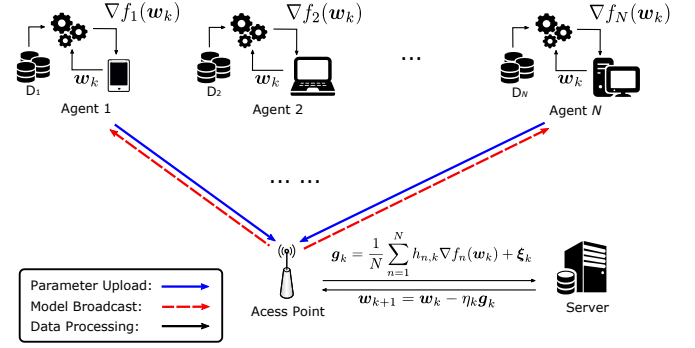}}

  \caption{ An illustration of the model training process: (A) agents evaluate local gradients based on their own dataset and upload to the server via analog transmissions, (B) the server receives the aggregated noisy gradient and use it to update the global model, (C) the new model is sent back to the agents, and the process is repeated. }
  \label{fig:FL_WN}
\end{figure}

\subsection{Analog Over-the-Air Model Training}
Let $\boldsymbol{w}_k$ be the global model broadcasted by the server at communication round $k$. 
Owing to the high transmit power of the edge server, we assume the global model can be successfully received by all the agents. 
Then, each client $n$ calculates its gradient $\nabla f_n( \boldsymbol{w}_k )$ and constructs the following analog signal:
\begin{align} \label{equ:AnagMod}
x_n(t) = \langle \mathbf{s}(t), \nabla f_n( \boldsymbol{w}_k  ) \rangle
\end{align}
where $\langle \cdot, \cdot \rangle$ denotes the inner product between two vectors and $\mathbf{s}(t) = (s_1(t), s_2(t), ..., s_d(t))$, $0 < t <T$, is a set of orthonormal baseband waveforms that satisfies:
\begin{align}
&\int_0^T s^2_i(t) dt = 1,~~~ i = 1, 2, ..., d \\
&\int_0^T s_i(t) s_j(t) = 0, ~~~\text{if}~ i \neq j.
\end{align}
According to \eqref{equ:AnagMod}, the signal $x_n(t)$ is essentially a superposition of the analog waveforms whereas the magnitude of $s_i(t)$ equals to the $i$-th element of $\nabla f_n( \boldsymbol{w}_k  )$.\footnote{Note that the magnitude of the waveforms can also be set at the quantized values of the gradients to reduce implementation complexity.}

Once the transmit waveforms $\{x_n(t)\}_{n=1}^N$ have been assembled, the agents send them out concurrently into the spectrum. 
And the signal received at the edge server can be expressed as follows: 
\begin{align}
y(t) = \sum_{ n=1 }^N h_{n, t} P_{n} x_n(t) + \xi (t)
\end{align}
where $h_{n, t}$ is the channel fading experienced by agent $n$, $P_{n}$ stands for the corresponding transmit power, and $\xi(t)$ represents the interference. 
Without loss of generality, we assume the channel fading is independent and identically distributed (i.i.d.) across the agents and communication rounds, with mean $\mu$ and variance $\sigma^2$. 
And the transmit power is set to compensate for the large-scale path loss. 
In order to characterize the heavy-tailed nature of wireless interference, we consider $\xi(t)$ follows a symmetric $\alpha$-stable distribution. 
The properties of this distribution will be elaborated in the next section.

The received signal $y(t)$ will be past to a set of matched filters, where each of them is tuned as $s_i(t)$, and output the following vector:
\begin{align} \label{equ:AnlgGrdnt}
\boldsymbol{g}_k = \frac{1}{N} \sum_{n=1}^N h_{n, k} \nabla f_n ( \boldsymbol{w}_k ) + \boldsymbol{\xi}_k
\end{align}
where $\boldsymbol{\xi}_k$ is a $d$-dimensional random vector with each entry being i.i.d. and following an $\alpha$-stable distribution. 
The server then updates global parameter as follows
\begin{align} \label{equ:ParUpdt}
\boldsymbol{w}_{k+1} = \boldsymbol{w}_{k} - \eta_k \boldsymbol{g}_k
\end{align}
where $\eta_k$ is the learning rate. 

\remark{
    \textit{The analog over-the-air gradient aggregation boasts two unique advantages: (a) high spectral utilization, as the agents do not need to vie for radio access but can simultaneously upload their local parameters to the server, and (b) low hardware cost, as the agents do not need correct the channel gain and hence they can transmit at a relatively constant power level. }
}

\remark{
    \textit{Thanks to the randomness from channel fading and interference, the gradient information, $\nabla f_n(\boldsymbol{w}_k)$, of each agent $n$ is concealed inside the noisy aggregated gradient $\boldsymbol{g}_k$. As pointed out by \cite{ElgParIss:21}, this form provides inherent privacy protection.  }
}

\remark{
    \textit{The momentum algorithm \cite{AbaBarChe:16} can be easily integrated into the analog over-the-air GD to rev up the learning process. Particularly, the overall training procedure is identical to that presented in Section II-B, except for the global model update stage \eqref{equ:ParUpdt}, where instead of directly using the gradient, the global parameter is updated as follows:
\begin{align} \label{equ:exps_vt}
\boldsymbol{v}_k &= \beta \boldsymbol{v}_{k-1} + \boldsymbol{g}_k,\\ \label{equ:MmntWtp1}
\boldsymbol{w}_{k+1} &= \boldsymbol{w}_k - \eta_k \boldsymbol{v}_k
\end{align}
in which $\beta \in (0, 1)$ denotes the controlling factor, also known as the momentum weight. }
}

\subsection{Heavy-Tailed Interference}
The spectrum is by nature a shared medium. Therefore, signals sent over the wireless channels inevitably suffer interference from the other concurrent transmitters. 
And it has been amply demonstrated from both theoretical \cite{Mid:77} and empirical \cite{ClaPedRod:20} perspectives that electromagnetic interference generally obeys a heavy-tailed distribution. 
In that respect, we adopt the symmetric $\alpha$-stable distribution to model the statistics of interference $\xi(t)$. 
\begin{definition}
\textit{
    The random variable $\xi(t)$ is said to follow a symmetric $\alpha$-stable distribution if its characteristic function takes the following form: 
    \begin{align}
    \mathbb{E}\left[ e^{j \omega \xi(t) } \right] = \exp( - \delta^\alpha \vert \omega \vert^\alpha )
    \end{align}
    where $\delta>0$ and $\alpha \in (0, 2]$. The parameters $\delta$ and $\alpha$ are commonly known as the scale parameter and tail index, respectively.  
}
\end{definition}

It is worthwhile to note that $\alpha$-stable distributions do not possess an explicit form of the probability density function in general, aside from two special cases, i.e., if $\alpha = 1$, the distribution reduces to Cauchy and when $\alpha = 2$, it reduces to Gaussian.
The tail index $\alpha$ determines the heaviness of tail in the probability density function of $\xi(t)$. 
Particularly, as depicted in Fig.~\ref{fig:HeavTailSign}, smaller the $\alpha$, thicker the tail in the distribution, which implies the random variable $\xi(t)$ has a higher chance to attain a very large value.{\footnote{This statement can also be theoretically corroborated by noticing at any time instance $t$, the probability that the magnitude of the analog signal exceeds a large value is given by $\mathbb{P}(\vert \sin(t) + \xi(t) \vert > \lambda) \sim \lambda^{-\alpha}$, where $\lambda \gg 1$ \cite{SamTaq:17}. And it implies that a smaller tail index  of $\xi(t)$ results in a higher probability of observing severe fluctuations in the analog signal magnitude. }} 

\begin{figure}[t!]
  \centering

  \subfigure[\label{fig:1a}]{\includegraphics[width=0.95\columnwidth]{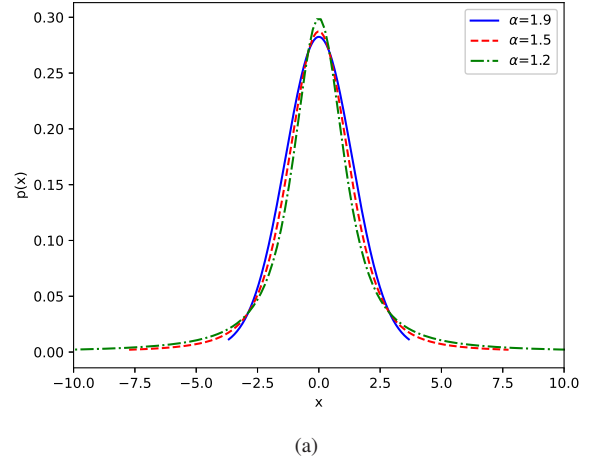}} 
  \subfigure[\label{fig:1b}]{\includegraphics[width=0.95\columnwidth]{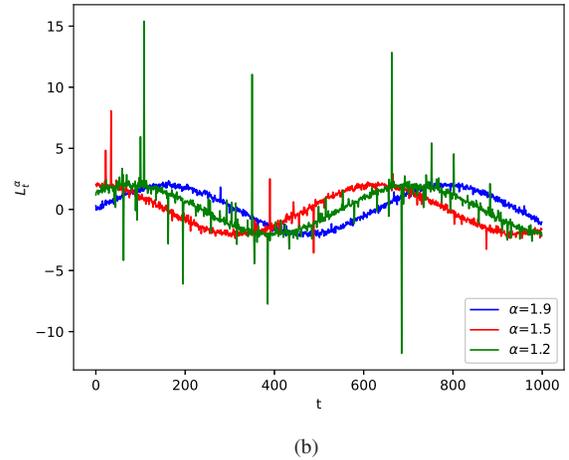}}
  \caption{ The examples of the $\alpha$-stable random variables. Fig.~2 (a) plots the probability density function under different values of the tail index. Fig.~2 (b) illustrates the effects of heavy-tailed noise imposed on three different sinusoid signals. }
  \label{fig:HeavTailSign}
\end{figure}

Actually, $\alpha$-stable random variables $\xi(t)$ have their moments being finite only up to the order $\alpha$, i.e., $\mathbb{E}[ \vert \xi(t) \vert^p ] < \infty$ only in the range of $0 \leq p \leq \alpha$. 
In this work, we consider $1 < \alpha < 2$, i.e., the interference $\xi(t)$ has zero mean but infinite variance. 
This, together with the random channel fading, makes the aggregated gradient $\boldsymbol{g}_k$ severely distorted. 
A natural question then arises as: \textit{Does analog over-the-air machine learning algorithms converge?}

We give an affirmative answer in the next section.

\section{Convergence Analysis }\label{sec:Analysis}
This section constitutes the main technical part of our paper, in which we derive analytical expressions for the convergence rate of the model training algorithm. 
For better readability, most proofs and mathematical derivations have been relegated to the Appendix.

\subsection{Preliminaries}
Because the second moment of the aggregated gradient does not exist, we opt for the $\alpha$ moment as an alternative. 
Correspondingly, we need the concepts of \textit{signed power} and \textit{$\alpha$-positive definite matrix} \cite{WanGurZhu:21}, respectively, for further mathematical manipulations. 

\begin{definition}
\textit{For a vector $\boldsymbol{w} = (w_1, ..., w_d)^{\mathsf{T}} \in \mathbb{R}^d$, we define its signed power as follows
\begin{align}
\boldsymbol{w}^{\langle \alpha \rangle } = \left( \mathrm{sgn}(w_1) \vert w_1 \vert^\alpha, ...,  \mathrm{sgn}(w_d) \vert w_d \vert^\alpha \right)^{\mathsf{T}}
\end{align}
where $\mathrm{sgn}(x) \in \{ -1, +1 \}$ takes the sign of the variable $x$. }
\end{definition}

\begin{definition}
\textit{A symmetric matrix $\boldsymbol{Q}$ is said to be $\alpha$-positive definite if $\langle \boldsymbol{v}, \boldsymbol{Q} \boldsymbol{v}^{\langle \alpha - 1 \rangle} \rangle > 0$ for all $\boldsymbol{v} \in \mathbb{R}^d$ with $\Vert \boldsymbol{v} \Vert_\alpha > 1$.    
}
\end{definition}

To facilitate the analysis, we make the following assumptions.

\begin{assumption}
\textit{The objective function $f: \mathbb{R}^d \rightarrow \mathbb{R}$ is $\gamma$-strongly convex, i.e., for any $\boldsymbol{w}, \boldsymbol{v} \in \mathbb{R}^d$ it is satisfied: 
    \begin{align}
    f( \boldsymbol{w} ) \geq f( \boldsymbol{v} ) + \langle \nabla f( \boldsymbol{v} ), \boldsymbol{w} - \boldsymbol{v} \rangle + \frac{\gamma}{2} \Vert \boldsymbol{w} - \boldsymbol{v} \Vert^2.
    \end{align}    
}
\end{assumption}

\begin{assumption}
\textit{The objective function $f: \mathbb{R}^d \rightarrow \mathbb{R}$ is $\lambda$-smooth, i.e., for any $\boldsymbol{w}, \boldsymbol{v} \in \mathbb{R}^d$ it is satisfied: 
    \begin{align}
    f( \boldsymbol{w} ) \leq f( \boldsymbol{v} ) + \langle \nabla f( \boldsymbol{v} ), \boldsymbol{w} - \boldsymbol{v} \rangle + \frac{\lambda}{2} \Vert \boldsymbol{w} - \boldsymbol{v} \Vert^2.
    \end{align}    
}
\end{assumption}

\begin{assumption}
\textit{For any given vector $\boldsymbol{w} \in \mathbb{R}^d$, the Hessian matrix of $f(\boldsymbol{w})$, i.e., $\nabla^2 f(\boldsymbol{w})$, is $\alpha$-positive definite. }
\end{assumption}


Since the magnitude of the transmitted waveforms cannot be arbitrarily large, we further assume the gradients of each agent is bounded, i.e., $\Vert \nabla f_n(\boldsymbol{w}) \Vert \leq G$, $\forall n \in \{1, ..., N\}$. 
Furthermore, because each element of $\boldsymbol{\xi}_k$ has a finite $\alpha$ moment, we consider the $\alpha$ moment of $\boldsymbol{\xi}_k$ is upper bounded by a constant $C$, i.e., $\mathbb{E}[ \Vert \boldsymbol{\xi}_k \Vert_\alpha^\alpha ] \leq C$.

\subsection{Convergence Rate of Analog Over-the-Air GD}
Armed with the notion of signed power, we can establish a Taylor expansion type inequality for the $\alpha$ norm vectors. 
\begin{lemma}
\textit{Given $\alpha \in [1, 2]$, for any $\boldsymbol{w}, \boldsymbol{v} \in \mathbb{R}^d$, the following holds:
\begin{align}
\Vert \boldsymbol{w} + \boldsymbol{v} \Vert_\alpha^\alpha \leq \Vert \boldsymbol{w} \Vert_\alpha^\alpha + \alpha \langle \boldsymbol{w}^{\langle \alpha - 1 \rangle }, \boldsymbol{v} \rangle + 4 \Vert \boldsymbol{v} \Vert_\alpha^\alpha.
\end{align}
}
\end{lemma}
\begin{IEEEproof}
Please refer to \cite{Kar:69}.
\end{IEEEproof}

Additionally, we lay out two technical lemmata that we would use to prove the main results. 
\begin{lemma}
\textit{Let $\boldsymbol{Q}$ be an $\alpha$-positive definite matrix, for $\alpha \in [1, 2]$, there exists $\kappa, L >0$, such that
\begin{align}
\Vert \boldsymbol{I} - k \boldsymbol{Q} \Vert_\alpha^\alpha \leq 1 - L k, \qquad \quad \forall k \in [0, \kappa)
\end{align}
 }
\end{lemma}
\begin{IEEEproof}
Please see Theorem~10 of \cite{WanGurZhu:21}. 
\end{IEEEproof}

\begin{lemma}
\textit{For a sequence of real numbers $\{ b_k \}$, $k \geq 1$, that satisfies:
\begin{align}
b_{k+1} \leq \left( 1 - \frac{c}{k} \right) b_k + \frac{c_1}{k^{p+1}}
\end{align}
where $c > p> 0$, $c_1 > 0$. The following relationship holds
\begin{align}
b_k \leq \frac{ c_1 }{ c - p } \cdot \frac{1}{k^p} + o\left( \frac{1}{k^{p+1}} + \frac{1}{k^c} \right)
\end{align}
where $o(\cdot)$ is the ``little o'' notation, meaning that if $f(k) = o(g(k))$ then $\forall B > 0$, there exists $k_0$ such that $f(k) \leq B g(k)$ for all $k \geq k_0$.
}
\end{lemma}
\begin{IEEEproof}
Please see Lemma~1 of \cite{Chung:54}. 
\end{IEEEproof}

We are now in position to present the main theoretical finding of this paper. 

\begin{theorem} \label{thm:ConvAnals}
\textit{Under the employed edge learning system, if the learning rate is set as $\eta_k = \theta/k$ where $\theta > \frac{ \alpha - 1 }{ \mu L }$, then the algorithm converges as
    \begin{align} \label{equ:ConvRt_OAML}
    \mathbb{E}\big[ \Vert \boldsymbol{w}_k - \boldsymbol{w}^* \Vert_{\alpha}^\alpha \big] \leq \frac{ 4 \theta^\alpha \Big( C + \frac{ \sigma^\alpha G^\alpha d^{ 1 - \frac{1}{\alpha} } }{ N^{\alpha/2} } \Big) }{ \mu \theta L - \alpha + 1 } \cdot \frac{1}{ k^{\alpha - 1} }.
    \end{align}
}
\end{theorem}
\begin{IEEEproof}
See Appendix~\ref{Apndx:ConvAnals_proof}. 
\end{IEEEproof}

A number of remarks are immediately in order. 

\remark{
    \textit{Although the aggregated gradient is corrupted by random channel fading and heavy-tailed interference with infinite variance, GD-based model training with a diminishing learning rate can converge to the global optimum without any modification, neither to the loss function nor the algorithm itself. In other words, GD-based algorithms are resilient to parameter distortions and hence are particularly suitable for edge learning systems.
    }
}

\remark{
    \textit{The convergence rate is the order of $\mathcal{O}( \frac{1}{ k^{\alpha - 1} } )$, which is dominated by the tail index $\alpha$. Specifically, a small $\alpha$ leads to a heavy tail in the distribution of interference, and that results in a slow convergence of the learning algorithm. }
}

\remark{
    \textit{The tail index, $\alpha$, also has an influence on the multiplication term in the convergence rate. Particularly, when $\mu L + 1> \alpha $, we can set $\theta = 1$ and because $N$ is usually a large number, it is safe to assume $\sigma G/N <1$. Then, from \eqref{equ:ConvRt_OAML} we can see that a decrease in $\alpha$ increases the multiplier term, which results in a slow convergence rate. }
}

\remark{
    \textit{If the variance of channel fading, $\sigma$, goes up, the channel will have a higher chance to encounter deep fade which inflicts the model training process. The effect is reflected in the multiplier of the convergence rate. }
}

\remark{
    \textit{An increase in the number of agents, $N$, can mitigate the impact of channel fading and accelerate the convergence rate. Therefore, scaling up the system can be beneficial to the federated learning. This is in line with conclusions made in \cite{SerCoh:20TSP}. Nonetheless, even if the channel fadings vanished, the convergence rate is still determined by the interference. }
}

Notably, the learning rate is also amenable, and the convergence rate can be characterized accordingly. 
\begin{corollary} \label{cor:ConvAnals}
\textit{Under the employed edge learning system, if the learning rate is set as $\eta_k \asymp k^{-\rho}$ with $\rho \in (0, 1)$, the training algorithm converges as 
\begin{align}
\mathbb{E}\big[ \Vert \boldsymbol{w}_k - \boldsymbol{w}^* \Vert_{\alpha}^\alpha \big] = \mathcal{O}\left( k^{ - \rho (\alpha - 1) } \right).
\end{align}
}
\end{corollary}
\begin{IEEEproof}
See Appendix~\ref{Apndx:Stp_ConvAnals_proof}. 
\end{IEEEproof}
From Corollary 1 we can see that the learning rate has a direct impact on the convergence property of the algorithm. 
Specifically, reducing the learning rate, as we decreases $\rho$, leads to a slowdown in convergence of the algorithm. 
In fact, the analog over-the-air GD converges even for very slowly decaying learning rate with $\rho$ being close to 0.

\subsection{Convergence Rate of Analog Over-the-Air Momentum}
In this part, we characterize the convergence rate of analog over-the-air momentum algorithm. To begin with, let us rewrite the intermediate parameter $\boldsymbol{v}_{k}$ in another form.
\begin{lemma}
\textit{
    The parameter $\boldsymbol{v}_k$ can be equivalently written as follows:
    \begin{align} \label{equ:expnd_vt}
    \boldsymbol{v}_k = \sum_{ i=1 }^{k} \beta^{k-i} \boldsymbol{g}_i.
    \end{align}
}
\end{lemma}
\begin{IEEEproof}
This result is obtained by recursively expanding $\boldsymbol{v}_i$ in terms of $\boldsymbol{g}_i$, $i = 1, ..., k$, according to \eqref{equ:exps_vt}.
\end{IEEEproof}

Following \eqref{equ:expnd_vt}, we can see that $\boldsymbol{v}_k$ is a moving average of the past noisy gradients, whereas the older ones are assigned with smaller weights.
The intuition behind the above operation is to add a ``heavy ball'', i.e., $\boldsymbol{v}_{k}$, in the update of parameters so as to alleviate the oscillations along the update path and bolster faster convergence.
On the basis of this lemma, we can further derive the convergence rate of the analog over-the-air algorithm as follows. 
\begin{theorem} \label{thm:ConvAnalsMnt}
\textit{When the employed edge learning system adopts momentum, if the learning rate is set as $\eta_k = \theta/k$ where $\theta > \frac{ (\alpha - 1)( 1 - \beta ) }{ \mu L }$, then the algorithm converges as
    \begin{align} \label{equ:ConvRt_OAMLM}
    \mathbb{E}\big[ \Vert \boldsymbol{w}_k - \boldsymbol{w}^* \Vert_{\alpha}^\alpha \big] \leq \frac{ 4 \theta^\alpha \Big( \frac{ 4C }{ 1 - \beta^\alpha } + \frac{ \sigma^\alpha G^\alpha d^{ 1 - \frac{1}{\alpha} } }{ N^{\frac{\alpha}{2}} ( 1 - \beta^2 )^{\frac{\alpha}{2}} } \Big) }{ \frac{ \mu \theta L }{ 1 - \beta }  - \alpha + 1 } \cdot \frac{1}{ k^{\alpha - 1} }.
    \end{align}
}
\end{theorem}
\begin{IEEEproof}
See Appendix~\ref{Apndx:ConvAnalsMnt_proof}. 
\end{IEEEproof}

We highlight two important observations from Theorem~2. 

\remark{
    \textit{Analog over-the-air momentum is bound to converge in the presence of heavy-tailed interference, and the convergence rate is $\mathcal{O}( \frac{1}{ k^{\alpha - 1} } )$ which is in the same order as analog over-the-air GD. }
}

\remark{
    \textit{The momentum weight, $\beta$, affects the multiplication term in the convergence rate, and it shall be adequately adjusted so as to attain fast convergence of the algorithm. }
}

Besides, analog over-the-air momentum also converges under a slower learning rate. The corresponding convergence rate is given by the following.
\begin{corollary} \label{cor:ConvAnalsMmnt}
\textit{When the employed edge learning system adopts momentum, if the learning rate is set as $\eta_k \asymp k^{-\rho}$ with $\rho \in (0, 1)$, the training algorithm converges as 
\begin{align}
\mathbb{E}\big[ \Vert \boldsymbol{w}_k - \boldsymbol{w}^* \Vert_{\alpha}^\alpha \big] = \mathcal{O}\left( k^{ - \rho (\alpha - 1) } \right).
\end{align}
}
\end{corollary}

From above discussions we can conclude that for over-the-air model training, both GD and momentum based methods converge in the same order. But one can control the momentum weight to reduce the multiplication term and speedup the algorithm.

\section{Generalization Error }\label{sec:Generalization}
This section presents the generalization capability of the trained model, which is quantified via the \textit{generalization error}.
Specifically, let $\mathcal{Z} = \mathcal{X} \times \mathcal{Y}$ denote the space of data points, where $\mathcal{X}$ and $\mathcal{Y}$ stand for the spaces of features and labels, respectively.
Then, the population risk is given by 
\begin{align}
\mathcal{R}( \boldsymbol{w} ) = \mathbb{E}_{\mathcal{Z}} \left[ \ell\left( \boldsymbol{w}; \boldsymbol{x}, {y} \right) \right]
\end{align}
and the empirical risk is 
\begin{align}
\hat{ \mathcal{R} } ( \boldsymbol{w}, \mathcal{D} ) = \frac{ 1 }{ \vert \mathcal{D} \vert } \sum_{ i=1 }^{ \vert \mathcal{D} \vert } \ell( \boldsymbol{w}; \boldsymbol{x}_i, {y}_i )
\end{align}
where $\mathcal{D} = \mathcal{D}_1 \cup \cdots \cup \mathcal{D}_n$ is the aggregated dataset. 
Suppose the learning algorithm has been executed for $K$ rounds of global iterations, the generalization error of the trained model is defined as 
\begin{align}
\mathcal{I}( \boldsymbol{w}_K ) = \left\vert \mathcal{R}( \boldsymbol{w}_K ) - \hat{ \mathcal{R} } ( \boldsymbol{w}_K, \mathcal{D} ) \right\vert,
\end{align}
which represents the expected difference between the error a model incurs on a training set versus the error incurred on a new data point.

In what follows, we elaborate on the steps toward a thorough analysis of the generalization error of the analog over-the-air model training. 

\subsection{Preliminaries}
Following \eqref{equ:ParUpdt}, during a specific communication round, the update of global parameter at the edge server can be written as follows:
\begin{align} \label{equ:A-OTA-SGD-R}
&\boldsymbol{w}_{k+1} - \boldsymbol{w}_k 
\nonumber\\
&= - \eta_k \mu \nabla f( \boldsymbol{w}_k ) - \frac{ \eta_k }{N} \! \sum_{n=1}^N \! \left( h_{n,k} - \mu \right) \! \nabla f_{n}( \boldsymbol{w}_k ) - \eta_k \boldsymbol{\xi}_k
\nonumber\\
&\stackrel{(a)}{ \approx } - \eta_k \mu \nabla f( \boldsymbol{w}_k ) - \Psi( \boldsymbol{w}_k ) \boldsymbol{n}_k - \eta_k \boldsymbol{\xi}_k
\end{align}
in which $\Psi( \boldsymbol{w}_k )$ is a $d \times d$ symmetric positive semi-definite matrix and $\boldsymbol{n}_k$ is a $d$ dimensional vector where each entry follows a independent normal distribution with zero mean and unit variance. 
It is noteworthy that the approximation ($a$) results from the fact that $\{ h_{n, k} \}_{n=1}^N$ are i.i.d. and by using the central limit theorem. 

As the number of communication rounds, $k$, becomes large, we have $\eta_k \ll 1$, and the recursion \eqref{equ:A-OTA-SGD-R} can be regarded as the discretization of a continuous time Feller process, which has the following form
\begin{align} \label{equ:FelProc}
d \mathbf{w}_t = - \nabla f( \mathbf{w}_t ) dt + \Psi( \mathbf{w}_t ) d \mathbf{B}_t + d \mathbf{L}_t
\end{align}
where $\mathbf{B}_t$ is a $d$ dimensional Brownian motion and $\mathbf{L}_t$ is a $d$ dimensional Levy process. 
In other words, the trajectory of the training process can be regarded as the composition of a drift, which is dominated by the gradient $\nabla f(\boldsymbol{w})$, a Brownian motion with a state-dependent covariance matrix, and an independent Levy process induced by the interference. A formal definition of Levy process is given as follows. 
\begin{definition}
\textit{ A Levy process $\{ \mathbf{L}_t \}_{ t \geq 0 }$ with the initial point $\mathbf{L}_0 = \mathbf{0}$ is defined by the following properties:
\begin{itemize}
    \item[(i)] Fix an arbitrary sequence $t_0 < t_1 < ... < t_k < ...$, the increment $\mathbf{L}_{t_i} - \mathbf{L}_{t_{i-1}}$ is independent for all $i$.
    \item[(ii)] For any $t > s > 0$, the quantities $\mathbf{L}_t - \mathbf{L}_s$ and $\mathbf{L}_{t-s}$ have the same distribution.
    \item[(iii)] $\mathbf{L}_t$ is continuous in probability, i.e., for any $\varepsilon$ and $s > 0$, there is $\mathbb{P}( \vert \mathbf{L}_t - \mathbf{L}_s \vert ) \rightarrow 0$ as $t \rightarrow s$.
\end{itemize}
 }
\end{definition}
In this work, since the increments of the Levy process follows a symmetric $\alpha$-stable process, it can be also termed as the symmetric $\alpha$-stable process. 

The Feller process $\mathbf{w}_t$ in \eqref{equ:FelProc} can be fully described by its characteristic exponent, given by 
\begin{align}
\varphi(\boldsymbol{w}, \boldsymbol{u} ) &= - \frac{ \log \left( \mathbb{E} \left[ e^{ - j \langle \boldsymbol{u}, \mathbf{w}_t \rangle } \right] \right)  }{ t }  
\nonumber\\&=   j \langle - \nabla f( \boldsymbol{w}), \boldsymbol{u} \rangle + \frac{1}{2} \langle \boldsymbol{u}, \Psi( \boldsymbol{w} ) \boldsymbol{u} \rangle 
\nonumber\\
&\qquad ~~~\,  + \int_{ \mathbb{R}^d \setminus \{ 0\} } \! \Big( 1 - e^{ j \langle \boldsymbol{u}, \boldsymbol{z} \rangle } + \frac{ j \langle \boldsymbol{u}, \boldsymbol{z} \rangle }{ 1 + \Vert \boldsymbol{z} \Vert^2 } \Big) \nu \! \left(d \boldsymbol{z} \right) 
\end{align}
where $j = \sqrt{-1}$, $\boldsymbol{w}, \boldsymbol{z}, \boldsymbol{u} \in \mathbb{R}^d$, and $\nu( \cdot )$ is a Levy measure.
Moreover, because the Levy process $\mathbf{L}_t$ is independent from the other components, the Feller process $\mathbf{w}_t$ is decomposible, namely, there exists $\boldsymbol{w}_0$ such that $\varphi(\boldsymbol{w}, \boldsymbol{u} ) = \varphi( \boldsymbol{u} ) + \tilde{\varphi}(\boldsymbol{w}, \boldsymbol{u} )$ where $\varphi( \boldsymbol{u} ) = \varphi(\boldsymbol{w}_0, \boldsymbol{u} )$ and $\tilde{\varphi}(\boldsymbol{w}, \boldsymbol{u} ) = \varphi(\boldsymbol{w}, \boldsymbol{u} ) - \varphi(\boldsymbol{w}_0, \boldsymbol{u} )$.

\subsection{Analysis of Generalization Error}
Let us denote the iterative training algorithm in \eqref{equ:ParUpdt} as $\mathcal{A}$, which is affected by two variables, i.e., the dataset $\mathcal{D}$ and the randomness $U$ which arised from the channel fading and interference.
At any timestamp $t$, $\mathbf{w}_t = [ \mathcal{A}( \mathcal{D}, U ) ]_t$ is the parameter returned by algorithm $\mathcal{A}$. 
Without loss of generality, we confine $t$ to be in the range of $[0, 1]$. 
Then, with intakes $\mathcal{D}$ and $U$, the training algorithm $\mathcal{A}$ outputs a stochastic process $\{ \mathbf{w}_t \}_{ t \in [0, 1] }$ which is the trajectory of iteration updates. 
For any given dataset $\mathcal{D}$, we write the whole trajectory of the Feller process as $\mathbf{w}_{ \mathcal{D} } = \{ \mathbf{w}_t = [ \mathcal{A}( \mathcal{D}, U ) ]_t \}_{ t \in [0, 1] }$. 

The intrinsic complexity of a Feller process is typically characterized by the notion of \textit{Hausdorff dimension}, which is formally defined as follows. 
\begin{definition}
\textit{
    The Hausdorff dimension of a Borel set $E \subset \mathbb{R}^d$ is defined as 
    \begin{align}
    \mathrm{dim}_{\mathrm{H}} E &= \sup \left\{ s > 0: \Lambda^s(E) > 0 \right\} \\
    &= \inf \left\{ s > 0: \Lambda^s(E) < \infty \right\}
    \end{align}
    where $\Lambda^s(E)$ is the $s$-dimensional Hausdorff measure, defined by 
    \begin{align}
    \Lambda^s(E) = \lim_{\varepsilon \rightarrow 0} \Lambda_{\varepsilon}^s(E)
    \end{align}
    in which $\Lambda_{\varepsilon}^s(E)$ is given by 
    \begin{align}
    &\Lambda_{\varepsilon}^s(E) 
    \nonumber\\
    &= \inf \left\{ \sum_{ n = 1 }^\infty  \left( \mathrm{diam} \, E_n \right)^s \!: E \subset \cup_{n=1}^\infty E_n ~\text{and}~ \mathrm{diam} \, E_n \leq \varepsilon \right\}
    \end{align}
    where $\mathrm{diam}$ stands for the diameter of a set and the infimium is take over all the $s$-coverings of $E$.
}
\end{definition}

The Hausdorff measure quantifies, informally, the ``roughness'' of an object in $\mathbb{R}^d$. For Levy (or stable) processes, this metric is deeply connected to the tails: 
\begin{lemma}
\textit{
    Let $\boldsymbol{\mathcal{L}} = \{ \mathbf{L}_t \}_{ t \in [0, 1] }$ be a symmetric $\alpha$-stable process in $\mathbb{R}^d$, where $0 < \alpha \leq 2$. Then, we have
    \begin{align}
    \mathrm{dim}_{\mathrm{H}} \boldsymbol{\mathcal{L}} = \alpha.
    \end{align}
}
\end{lemma}
\begin{IEEEproof}
Please see Theorem~4 of \cite{BluGet:60}.
\end{IEEEproof}

Furthermore, because for any given dataset $\mathcal{D}$, the Feller process $\mathbf{w}_t$ resulted from algorithm $\mathcal{A}$ always pertains an independent Levy (or $\alpha$-stable, in our case) process. Then, we have the following result. 
\begin{lemma}
\textit{Let $\mathbf{w}_{\mathcal{D}}$ be the trajectory of the Feller process given by \eqref{equ:FelProc}. Then, we have 
\begin{align}
\mathrm{dim}_{ \mathrm{H} } \mathbf{w}_{\mathcal{D}} \leq \alpha = \inf \left\{ s \leq 0: \lim_{ \Vert \boldsymbol{u} \Vert \rightarrow \infty } \frac{ \vert \varphi ( \boldsymbol{w} ) \vert }{ \Vert \boldsymbol{u} \Vert^s } = 0 \right\}.
\end{align}
}
\end{lemma}
\begin{IEEEproof}
Please see Theorem~4 of \cite{Sch:98}.
\end{IEEEproof}

To facilitate the analysis of the generalization error of algorithm $\mathcal{A}$, we assume the loss function on each data point, $\ell$, is bounded by a constant $B$.\footnote{ The assumption on the boundedness of the loss function can in fact be relaxed, although that will give rise to additional complexities in the analysis. } 
To this end, we are ready to present the final result. The proof is broadly similar to the previous work \cite{SimSenDel:20}, albeit with some minor modifications.

\begin{theorem} \label{thm:GenErr}
\textit{When the number of data samples, $\vert \mathcal{D} \vert$, is sufficiently large, the following holds
\begin{align}
\sup_{ \boldsymbol{w} \in \mathbf{w}_{\mathcal{D}} } \left\vert \hat{\mathcal{R}}( \boldsymbol{w}, \mathcal{D} ) - \mathcal{R}( \boldsymbol{w} ) \right\vert \leq B \sqrt{ \frac{ 2 \alpha \log( \lambda^2 \vert \mathcal{D} \vert ) }{ \vert \mathcal{D} \vert } + \frac{ \log(1/p) }{ \vert \mathcal{D} \vert } }
\end{align} 
with probability at least $1-p$.
}
\end{theorem}
\begin{IEEEproof}
See Appendix~\ref{Apndx:GenErr_proof}.
\end{IEEEproof}
From this theorem, we can see that with a certain probability, the generalization error decreases with respect to the tail index. In other words, the interference bears the potential to improve the generalization capability of the trianed model.

\remark{
    \textit{In a similar spirit, one can approximate the training procedure of analog over-the-air GD with momentum by its continuous time proxy \cite{SimZhuTeh:20ICML} and arrive at conclusions as per Theorem~3.}
}

\section{ Simulation Results }\label{sec:NumResult}
In this section, we conduct experimental evaluation of the wireless machine learning algorithm.{\footnote{Although this paper concentrates on the theoretical understanding of the over-the-air aggregation for machine learning, such a scheme has been implemented in a real world prototype \cite{GuoZhuMa:21}. }}
Particularly, we examine the performance of the analog over-the-air GD for training a multi layer perceptron (MLP) on the MNIST dataset which contains the hand-written digits \cite{LeCBotBen:98}. 
The MLP is consisted of 2 hidden layers, each has 64 units and adopts the ReLu activations. 
We extract 60,000 data points from the MNIST dataset for training, where each agent is assigned with an independent portion that contains 600 data samples.
Furthermore, we adopt the Rayleigh fading to model the channel gain. 
Unless otherwise stated, the following parameters will be used: Learning rate exponent $\rho = 1$, tail index $\alpha = 1.5$, number of agents $N=100$, average channel gain $\mu = 1$. 
The experiments are implemented with Pytorch on Tesla P100 GPU and averaged over 4 trials.

In Fig.~\ref{fig:AlphaEffc}, we plot the training loss as a function of the communication rounds under different values of the tail index. 
We can see that the training loss declines steadily along with the communication rounds, regardless of the heaviness of the tail in the interference's distribution. 
Nonetheless, the tail index $\alpha$ still plays a vital role in the rate of convergence. 
Particularly, with an increase in the tail index, the convergence rate goes up accordingly, whereas the improvement is non-linear with respect to $\alpha$. 
These observations corroborate the statement of Theorem~1.  
\begin{figure}[t!]
  \centering{}

    {\includegraphics[width=0.95\columnwidth]{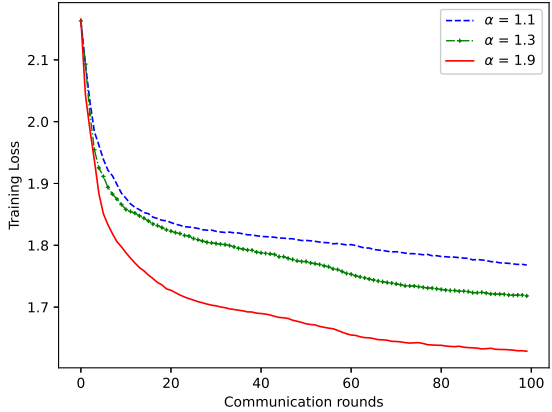}}

  \caption{ Simulation results of the training loss of MLP on the MNIST data set, under different tail index $\alpha$. }
  \label{fig:AlphaEffc}
\end{figure}

Next, we evaluate the effects of the number of participating agents on the model training procedure. 
It can be seen from Fig.~\ref{fig:PartNumEffc} that the convergence rate of analog over-the-air GD increases with $N$, revealing a positive influence from the enlarged number of agents in the system.
The reason attributes to two aspects: ($a$) as we fix the size of dataset per agent, an increase in $N$ boosts up the utilization of data information because all the agents can concurrently access the radio channel and participate in each round of global iteration, and ($b$) allowing more agents to partake in the analog transmission can reduce the impact of channel fading, as explained in Remark~8. 
Nevertheless, we also observe that such an effect is less significant compared to the tail index because it only influences the multiplier in the convergence rate.

\begin{figure}[t!]
  \centering{}

    {\includegraphics[width=0.95\columnwidth]{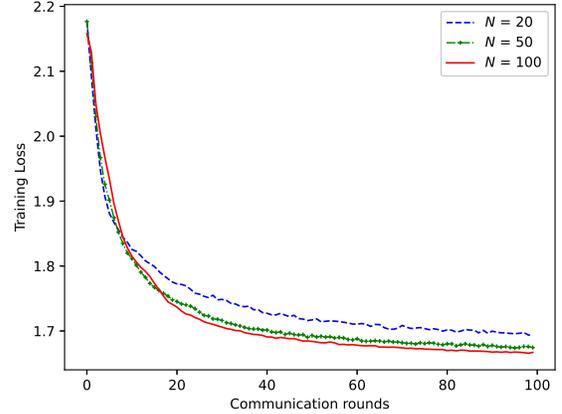}}

  \caption{ Simulation results of the training loss of MLP on the MNIST data set, under different number of agents $N$. }
  \label{fig:PartNumEffc}
\end{figure}

Fig.~\ref{fig:Rho_Impact} illustrates the impact of learning rate, which is controlled by the parameter $\rho$, on the convergence rate of the algorithm. 
The figure shows that a decrease in $\rho$, which slows down the learning rate, can impede the convergence of the model training.
Note that compared to the number of agents, $N$, the variants in the learning rate has a more pronounced impact on the training algorithm as it affects the exponential parameter in the convergence rate.

\begin{figure}[t!]
  \centering{}

    {\includegraphics[width=0.95\columnwidth]{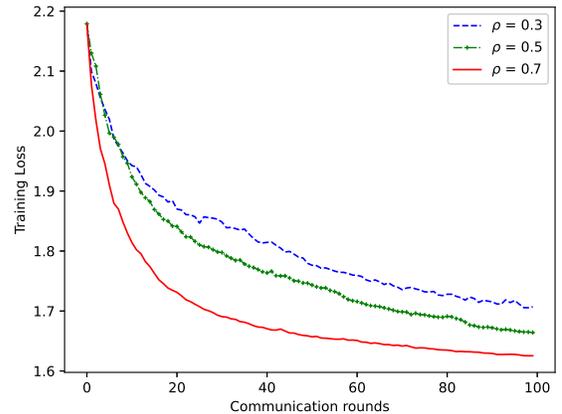}}

  \caption{ Simulation results for training on the MNIST data set, under different $\rho$. }
  \label{fig:Rho_Impact}
\end{figure}


We further demonstrate the effect of momentum in Fig.~\ref{fig:MomntEffc}, where the convergence curves are plotted under different momentum weights $\beta$.
First, we can see that all the loss functions decays gradually with the communication rounds, which validate that analog over-the-air GD can be run in conjunction with momentum. 
Second, by comparing the model training without momentum, i.e., $\beta=0$, and those with momentum weights set as $\beta = 0.05$ or $\beta = 0.2$, we observe that the momentum method is able to enhance the algorithm's convergence rate. 
Nonetheless, one shall also keep in mind that to reap such a benefit, the momentum weight, $\beta$, needs to be appropriately tuned. In fact, the comparison between the model convergence rate under $\beta = 0$ and $\beta = 0.1$ demonstrates that if the momentum weight is not set well, the scheme may even slow down the model training procedure. 
Notably, these observations also are aligned with the conclusions drawn in Remark~10. 
Finally, similar to the effects of $N$, momentum only affects the multiplier in the convergence rate and hence its variants do not lead to a severe fluctuation in the convergence rate. 

\begin{figure}[t!]
  \centering{}

    {\includegraphics[width=0.95\columnwidth]{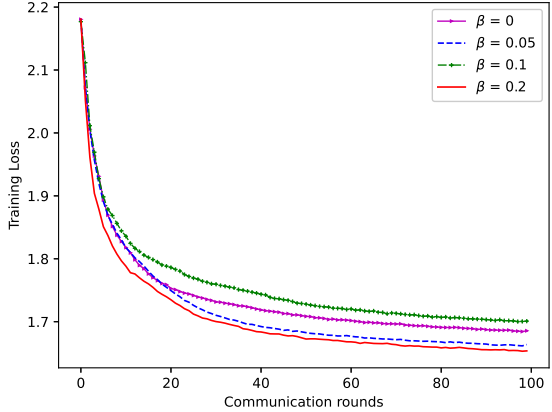}}

  \caption{ Simulation results of the training loss of MLP on the MNIST data set, under different momentum weights $\beta$. }
  \label{fig:MomntEffc}
\end{figure}


We now turn our attention to the generalization capability of the trained model. 
Because generalization of machine learning algorithms is characterized by its prediction accuracy on new data sets, we depict the test accuracy as a function of the communication rounds. 
For the task of training MLP on the MNIST dataset, we allocate 10,000 data points at the server for testing.
Besides, we add another experiment of learning a convolutional neural network (CNN) on the CIFAR-10 dataset \cite{KriHin:09}. The CIFAR-10 dataset consists of 60,000 colour images in 10 classes, with 6000 images per class. And the CNN has two convolutional layers with a combination of max pooling, followed by two fully-connected layers, then a softmax output layer. 
We extract 50,000 data points from the CIFAR-10 dataset for training, where each agent is assigned with an independent portion that contains 500 data samples.
At the server side, we allocate 10,000 data points for testing. 
The statistical models are trained via analog over-the-air GD and the testing dataset is used to evaluate the effectiveness of the trained results.
Moreover, we do not average over different trials in the following experiments since, as pointed out by Theorem~3, these events happen with a certain probability. 

Fig.~\ref{fig:AlphaAccEffc} illustrates the prediction accuracy of the machine learning algorithms under varying values of the tail index. 
From this figure, we can see that when the model training converges, the one that experienced a heavier tailed interference achieves a higher test accuracy. 
This implies that the heavy tail intrinsic of interference can improve the generalization capability of the model. 
The reason may be ascribed to the fact that the heavy-tailed interference sometimes incurs a ``big jump'' which largely deviates the global model. 
And such a deviation becomes beneficial if the global model is trapped in local minima.
Nonetheless, because the interference is independent from the state of model training, this gain from interference only occurs fortuitously.




\begin{figure}[t!]
  \centering

  \subfigure[\label{fig:7a}]{\includegraphics[width=0.95\columnwidth]{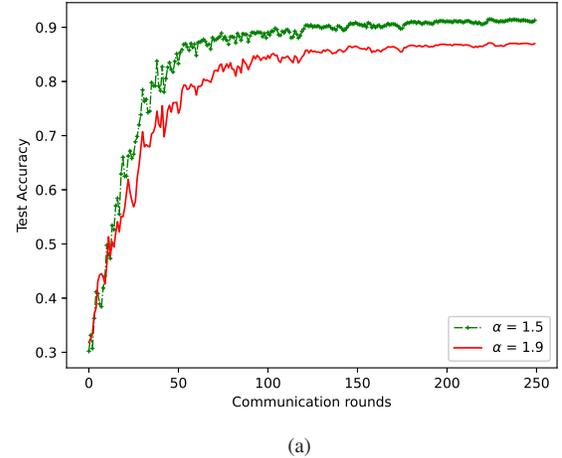}} ~
  \subfigure[\label{fig:7b}]{\includegraphics[width=0.95\columnwidth]{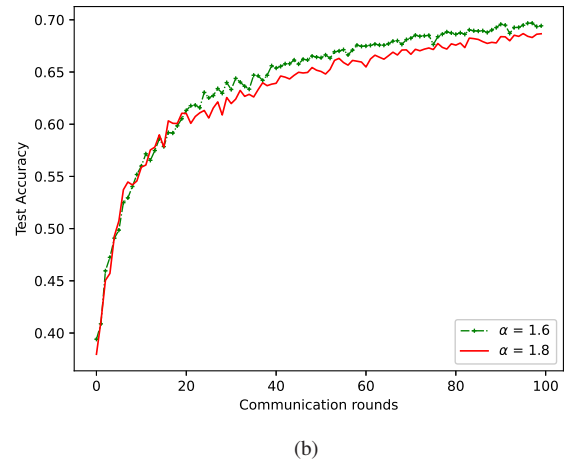}}
  \caption{ Simulation results of the test accuracy under analog over-the-air GD algorithm: ($a$) training the MLP on the MNIST dataset and ($b$) training the CNN on the CIFAR-10 dataset.  }
  \label{fig:AlphaAccEffc}
\end{figure}

The heavy tail's influence on generalization power is also evident in the analog over-the-air momentum training approach, as shown by Fig.~\ref{fig:AlphaAccEffc_Mntm}. 
To be more specific, in this experiment we conduct learning tasks as the aforementioned ones by training the models via analog over-the-air GD with momentum (cf. Remark~3) where the momentum weight is set as $\beta = 0.2$. 
And the figure discloses a similar phenomenon as the previoius one, that heavy tailed interference can enhance the generalization power of the trained model. 
This observation also validates our claim in Remark~11. 

\begin{figure}[t!]
  \centering

  \subfigure[\label{fig:8a}]{\includegraphics[width=0.95\columnwidth]{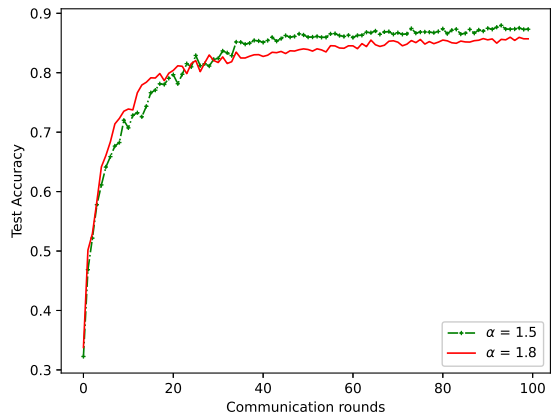}} ~
  \subfigure[\label{fig:8b}]{\includegraphics[width=0.95\columnwidth]{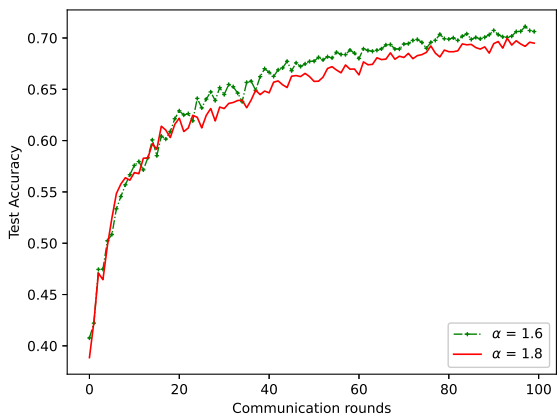}}
  \caption{ Simulation results of the test accuracy under analog over-the-air momentum algorithm: ($a$) training the MLP on the MNIST dataset and ($b$) training the CNN on the CIFAR-10 dataset.  }
  \label{fig:AlphaAccEffc_Mntm}
\end{figure}

\section{ Conclusion }\label{sec:Conclusion}
We have conducted a theoretical study of the analog over-the-air model training that solves a distributed machine learning problem in a wireless network. 
Specifically, the agents concurrently transmit an analog signal that is comprised of the local gradients. The edge server receives a superposition of the gradients, which are however distorted by the channel fading and interference, and uses this noisy gradient to update the global model. 
Due to the heavy-tailed intrinsic of interference, the aggregated gradient admits an infinite variance that hinders the use of conventional techniques for convergence analysis that relies on the existence of second moments.
To that end, we have taken a new route to establish the analyses of convergence rate and generalization error of the training algorithm.
Our analyses have unveiled a two-sided effect of the heavy-tailed interference on the overall learning procedure. 
More precisely, although heavy-tailed noise slows down the convergence rate, it has the potential to improve the generalization capability of the trained model.
Additionally, analog over-the-air training can be speeded up by adopting a momentum-like scheme.
These new results advance the understanding of wireless machine learning, and the techniques developed in this paper also provided an entryway toward the analysis of distributed learning algorithms in the context of heavy-tailed noise. 
A few possible extensions are enlisted below:

\begin{itemize}
    \item \textit{What about asynchronization in the transmissions?} In this paper, we implicitly assume the analog signals of the agents are synchronized in time. Such an assumption may not hold when the number of agents becomes very large. To investigate the effects of asynchronous transmissions on the learning algorithm, as well as approaches to alleviate this issue is an important direction. 
    \item \textit{How to stabilize the model training procedure?} Due to the heavy-tailed distribution, interference fluctuates abruptly and makes the model training less stable. To explore other algorithms for parameters updating that are less sensitive to heavy-tailed noise, e.g., via the gradient clipping \cite{GorDanGas:20}, can also be a good direction. 
    \item \textit{How about higher order methods?} The training algorithm we employed in this paper is based on GD, which is well known to be a first order scheme. Extending the framework to analyze analog over-the-air machine learning under GD-based variants including AdaGrad, AdaDelta, and Adam, or second order algorithms, such as ADMM and Quasi Newton method, is a concrete direction.
    \item \textit{Can we engineer the interference to facilitate wireless machine learning?} This paper shows that interference plays a pivotal role in wireless machine learning. As interference management has been a sophisticated research topic in conventional wireless communications, how to exploit methods such as multiple antenna transmissions or multi cell cooperations that can alter the statistics of interference to improve the performance of machine learning is another future direction.
\end{itemize}

\begin{appendix}
\subsection{Proof of Theorem~\ref{thm:ConvAnals}} \label{Apndx:ConvAnals_proof}
For ease of exposition, let us denote by $\Delta_k = \boldsymbol{w}_k - \boldsymbol{w}^*$. 
Then, in a specific communication round $k$, we can leverage \eqref{equ:AnagMod} and \eqref{equ:ParUpdt} to write the update of global parameter $\boldsymbol{w}_k$ as follows:
\begin{align} \label{equ:Delta_Updt}
& \big\Vert \Delta_{k+1} \big\Vert_\alpha^\alpha = \Big\Vert \boldsymbol{w}_k - \boldsymbol{w}^* - \eta_k \Big( \frac{1}{N} \sum_{n=1}^N h_{n, k} \nabla f_n ( \boldsymbol{w}_k ) + \boldsymbol{\xi}_k \Big) \Big\Vert_\alpha^\alpha 
\nonumber\\
&\stackrel{(a)}{\leq} \big\Vert \Delta_k -  \frac{\eta_k}{N} \sum_{n=1}^N h_{n, k} \nabla f_n ( \boldsymbol{w}_k ) \big\Vert_\alpha^\alpha + 4 \eta_k^\alpha \big\Vert \boldsymbol{\xi}_k \big\Vert_\alpha^\alpha
\nonumber\\
&+ \alpha \Big \langle \big( \Delta_k -  \frac{\eta_k}{N} \sum_{n=1}^N h_{n, k} \nabla f_n ( \boldsymbol{w}_k ) \big)^{\langle \alpha - 1 \rangle}, \eta_k \boldsymbol{\xi}_k \Big \rangle
\end{align}
where ($a$) follows from Lemma~1. 

By noticing that each entry of $\boldsymbol{\xi}_t$ is independent and has a zero mean, we can take an expectation on both sides of \eqref{equ:Delta_Updt} and arrive at the following:
\begin{align} \label{equ:Delta_Bound}
&\mathbb{E} \Big[ \big\Vert \Delta_{k+1} \big\Vert_\alpha^\alpha \Big] 
\nonumber\\
&\leq \mathbb{E} \Big[ \big\Vert \Delta_k -  \frac{\eta_k}{N} \sum_{n=1}^N h_{n, k} \nabla f_n ( \boldsymbol{w}_k ) \big\Vert_\alpha^\alpha \Big] + 4 \eta_k^\alpha \mathbb{E}\Big[ \big\Vert \boldsymbol{\xi}_k \big\Vert_\alpha^\alpha \Big]
\nonumber\\
&\leq \mathbb{E} \Big[ \big\Vert \Delta_k -  \frac{\eta_k}{N} \sum_{n=1}^N h_{n, k} \nabla f_n ( \boldsymbol{w}_k ) \big\Vert_\alpha^\alpha \Big] + 4 C \eta_k^\alpha. 
\end{align}
The first term on the right hand side of \eqref{equ:Delta_Bound} can be further bounded as follows:
\begin{align}
&\mathbb{E} \Big[ \big\Vert \Delta_k -  \frac{\eta_k}{N} \sum_{n=1}^N h_{n, k} \nabla f_n ( \boldsymbol{w}_k ) \big\Vert_\alpha^\alpha \Big]
\nonumber\\
&= \mathbb{E} \Big[ \big\Vert \Delta_k - \mu \eta_k \nabla f ( \boldsymbol{w}_k ) -  \frac{\eta_k}{N} \sum_{n=1}^N ( h_{n, k} - \mu ) \nabla f_n ( \boldsymbol{w}_k ) \big\Vert_\alpha^\alpha \Big]
\nonumber\\
&\leq \mathbb{E} \Big[ \big\Vert \Delta_k - \mu \eta_k \nabla f ( \boldsymbol{w}_k ) \big\Vert_\alpha^\alpha \Big] 
\nonumber\\
&+\! \alpha \mathbb{E}\bigg[ \! \Big\langle \! \Big(\! \Delta_k - \mu \eta_k \nabla f ( \boldsymbol{w}_k ) \! \Big)^{\!\langle \alpha - 1 \rangle}\!\!\!, \frac{\eta_k}{N} \! \sum_{n=1}^N ( h_{n, k} - \mu ) \nabla f_n ( \boldsymbol{w}_k ) \! \Big\rangle \! \bigg]
\nonumber\\
&\qquad \qquad \qquad \quad \qquad +  \frac{4 \eta^\alpha_k}{N^\alpha} \mathbb{E} \Big[ \big\Vert \sum_{n=1}^N ( h_{n, k} - \mu ) \nabla f_n ( \boldsymbol{w}_k ) \big\Vert_\alpha^\alpha \Big]
\nonumber\\
&\stackrel{(a)}{=} \underbrace{ \mathbb{E} \Big[ \big\Vert \Delta_k - \mu \eta_k \nabla f ( \boldsymbol{w}_k ) \big\Vert_\alpha^\alpha \Big] }_{Q_1}  
\nonumber\\
&\qquad \qquad \qquad \quad \qquad +  \frac{4 \eta^\alpha_k}{N^\alpha}  \underbrace{ \mathbb{E} \Big[ \big\Vert \sum_{n=1}^N ( h_{n, k} - \mu ) \nabla f_n ( \boldsymbol{w}_k ) \big\Vert_\alpha^\alpha \Big] }_{Q_2} 
\end{align}
where ($a$) follows from the fact that $\{ h_{n,k} \}_{n=1}^N$ are i.i.d. and satisfy $\mathbb{E}[ h_{n,k} ] = \mu$, $n = 1, ..., N$. 

For the empirical risk minimizer $\boldsymbol{w}^*$, we have $\nabla f(\boldsymbol{w}^*) = 0$. Hence $Q_1$ can be expanded as follows
\begin{align}
Q_1 &= \mathbb{E} \Big[ \big\Vert \Delta_k - \mu \eta_k \big[\, \nabla f ( \boldsymbol{w}_k ) - \nabla f ( \boldsymbol{w}^* ) \,\big] \big\Vert_\alpha^\alpha \Big]
\nonumber\\
&\stackrel{(a)}{=} \mathbb{E} \Big[ \big\Vert \Delta_k - \mu \eta_k \nabla^2 f ( \boldsymbol{w}^{\sharp}_k ) \Delta_k \big\Vert_\alpha^\alpha \Big]
\nonumber\\
&\leq \Big\Vert \boldsymbol{I}_d - \mu \eta_k \nabla^2 f ( \boldsymbol{w}^{\sharp}_k ) \Big\Vert_\alpha^\alpha \times \mathbb{E} \Big[ \big\Vert \Delta_k \big\Vert_\alpha^\alpha \Big]
\end{align}
where ($a$) follows from the mid value theorem, and $\boldsymbol{I}_d$ denotes a $d \times d$ identity matrix. Furthermore, by the property, we have 
\begin{align}
\Big\Vert \boldsymbol{I}_d - \mu \eta_k \nabla^2 f ( \boldsymbol{w}^{\sharp}_k ) \Big\Vert_\alpha^\alpha \leq \left( 1 - \eta_k \mu L \right).
\end{align}
Therefore, $Q_1$ can be bounded as 
\begin{align} \label{equ:Q_1_bnd}
Q_1 \leq \left( 1 - \eta_k \mu L \right) \mathbb{E} \Big[ \big\Vert \Delta_k \big\Vert_\alpha^\alpha \Big].
\end{align}
On the other hand, $Q_2$ can be bounded via the following:
\begin{align} \label{equ:Q_2_bnd}
&Q_2 = \mathbb{E} \Big[ \big\Vert \sum_{n=1}^N ( h_{n, k} - \mu ) \nabla f_n ( \boldsymbol{w}_k ) \big\Vert_\alpha^\alpha \Big]
\nonumber\\
&\stackrel{(a)}{\leq} d^{ 1 - \frac{1}{\alpha} } \cdot \mathbb{E} \Big[ \Big( \big\Vert \sum_{n=1}^N ( h_{n, k} - \mu ) \nabla f_n ( \boldsymbol{w}_k ) \big\Vert_2^2 \Big)^{ \frac{\alpha}{2} } \Big]
\nonumber\\
&\stackrel{(b)}{\leq} d^{ 1 - \frac{1}{\alpha} } \cdot  \Big( \mathbb{E} \Big[ \big\Vert \sum_{n=1}^N ( h_{n, k} - \mu ) \nabla f_n ( \boldsymbol{w}_k ) \big\Vert_2^2 \Big] \Big)^{ \frac{\alpha}{2} } 
\nonumber\\
&= d^{ 1 - \frac{1}{\alpha} } \! \cdot \!\! \Big(\! \mathbb{E} \Big[ \!\!\! \sum_{n, m=1}^N \!\!\!\! \big\langle ( h_{n, k} - \mu ) \nabla f_n ( \boldsymbol{w}_k ), ( h_{m, k} - \mu ) \nabla f_m ( \boldsymbol{w}_k ) \big\rangle \Big] \Big)^{ \frac{\alpha}{2} } 
\nonumber\\
&= d^{ 1 - \frac{1}{\alpha} } \cdot  \Big( \sum_{n=1}^N \mathbb{E} \Big[  ( h_{n, k} - \mu )^2 \cdot \big\Vert \nabla f_n ( \boldsymbol{w}_k ) \big\Vert_2^2 \Big] \Big)^{ \frac{\alpha}{2} } 
\nonumber\\
& \leq d^{ 1 - \frac{1}{\alpha} } \cdot \Big( N \sigma^2 G^2 \Big)^{ \frac{\alpha}{2} } = d^{ 1 - \frac{1}{\alpha} } \cdot  N^{ \frac{\alpha}{2} } \sigma^\alpha G^\alpha
\end{align}
where ($a$) and ($b$) follows from the Holder's inequality and Jensen's inequality, respectively. 
To this end, by substituting \eqref{equ:Q_1_bnd} and \eqref{equ:Q_2_bnd} into \eqref{equ:Delta_Bound}, we have 
\begin{align} \label{equ:DeltatFnalBnd}
&\mathbb{E} \Big[ \big\Vert \Delta_{k+1} \big\Vert_\alpha^\alpha \Big] \leq \left( 1 - \eta_k \mu L \right) \mathbb{E} \Big[ \big\Vert \Delta_k \big\Vert_\alpha^\alpha \Big] 
\nonumber\\
&\qquad \qquad \qquad \qquad \qquad \qquad \qquad + 4 \Big( C + \frac{ \sigma^\alpha G^\alpha d^{ 1 - \frac{1}{\alpha} } }{ N^{\alpha/2} } \Big) \eta_k^\alpha
\end{align}
and the proof is completed by invoking Lemma~2 and removing the higher order terms as they become infinitesmall as $k$ goes large.

\subsection{Proof of Corollary~\ref{cor:ConvAnals}} \label{Apndx:Stp_ConvAnals_proof}
In order to prove this corollary, let us first recall the following lemma.

\begin{lemma} (Lemma 4.2, \cite{Fab:67}) \label{lma:Apndx}
\textit{For a sequence of real numbers $\{ b_k \}$, $k \geq 1$, that satisfies the following recursion:
\begin{align}
b_{k+1} = \left( 1 - \frac{c}{k^p} \right) b_k + \frac{c_1}{k^{p+q}}
\end{align}
where $0 < p <1$ and $c, c_1, q > 0$. We have $b_k = \mathcal{O}(k^{-q})$. 
}
\end{lemma}

Then, if we assign $\eta_k = k^{-\rho}$ with $\rho \in (0, 1)$ in \eqref{equ:DeltatFnalBnd}, it yields 
\begin{align} 
&\mathbb{E} \Big[ \big\Vert \Delta_{k+1} \big\Vert_\alpha^\alpha \Big] \leq \left( 1 - \frac{\mu L}{k^\rho} \right) \mathbb{E} \Big[ \big\Vert \Delta_k \big\Vert_\alpha^\alpha \Big] 
\nonumber\\
&\qquad \qquad \qquad \qquad \qquad \qquad \quad~ + 4 \Big( C + \frac{ \sigma^\alpha G^\alpha d^{ 1 - \frac{1}{\alpha} } }{ N^{\alpha/2} } \Big)\frac{1}{k^{\alpha \rho}}.
\end{align}
And the result follows by invoking Lemma~\ref{lma:Apndx}.

\subsection{Proof of Theorem~\ref{thm:ConvAnalsMnt}} \label{Apndx:ConvAnalsMnt_proof}
Following similar lines in the proof of Theorem~1, we can expand, and bound, the progress of $\Delta_k$ in a specific communication round $k$ in the following way: 
\begin{align} \label{equ:Deltp1_Mmnt}
&\mathbb{E} \Big[ \big\Vert \Delta_{k+1} \big\Vert_\alpha^\alpha \Big] 
\nonumber\\
&\stackrel{(a)}{=}\mathbb{E} \Big[ \big\Vert \Delta_{k} - \eta_k \sum_{ i=1 }^{k} \beta^{k-i} \boldsymbol{g}_i \big\Vert_\alpha^\alpha \Big] 
\nonumber\\
&=\! \mathbb{E} \Big[ \big\Vert \Delta_k -\!  \frac{\eta_k}{N} \sum_{ i=1 }^{k} \beta^{k-i}  \sum_{n=1}^N h_{n, i} \nabla f_n ( \boldsymbol{w}_k ) - \eta_k \! \sum_{ i=1 }^{k} \beta^{k-i} \boldsymbol{\xi}_i \big\Vert_\alpha^\alpha \Big] 
\nonumber\\
&\stackrel{(b)}{\leq}\! \underbrace{\mathbb{E} \Big[ \big\Vert \Delta_k -\! \mu \eta_k \! \sum_{i=1}^k \beta^{k-i} \nabla f ( \boldsymbol{w}_i ) \big\Vert_\alpha^\alpha \Big]}_{Q_1} \!+ 4 \eta_k^\alpha \! \underbrace{\mathbb{E}\Big[ \big\Vert \sum_{ i=1 }^{k} \beta^{k-i} \boldsymbol{\xi}_i \big\Vert_\alpha^\alpha \Big]}_{Q_2}
\nonumber\\
&\qquad\qquad\quad + \frac{4 \eta_k^\alpha}{ N^\alpha } \underbrace{\mathbb{E} \Big[ \big\Vert \sum_{i=1}^k \beta^{k-i} \sum_{n=1}^N ( h_{n, i} - \mu ) \nabla f_n ( \boldsymbol{w}_k ) \big\Vert_\alpha^\alpha \Big]}_{Q_3}  
\end{align}
where ($a$) follows by using \eqref{equ:MmntWtp1} and \eqref{equ:expnd_vt}, while ($b$) by adopting Lemma~1 and the fact that the random elements $\boldsymbol{\xi}_i$ and $h_{n, i}$ are i.i.d. for $i = 1, ..., k$ and $n = 1, ..., N$, whereas $\mathbb{E}[\boldsymbol{\xi}_i] = 0$ and $\mathbb{E}[h_{n, i}] = \mu$.

Next, we bound $Q_1$, $Q_2$, and $Q_3$, respectively. First of all, by noticing that along the update path of $f(\boldsymbol{w})$, at the position $\boldsymbol{w}_k$, $\nabla f(\boldsymbol{w}_k)$ attains the steepest descent direction, we have 
\begin{align} \label{equ:BndQ1_Mmnt}
Q_1 &\leq \mathbb{E} \Big[ \big\Vert \Delta_k - \mu \eta_k \! \sum_{i=1}^k \beta^{t-i} \nabla f ( \boldsymbol{w}_k ) \big\Vert_\alpha^\alpha \Big]
\nonumber\\
&= \mathbb{E} \Big[ \big\Vert \Delta_k - \mu \eta_k \! \sum_{i=1}^k \beta^{k-i} \nabla f ( \boldsymbol{w}_k ) \big\Vert_\alpha^\alpha \Big]
\nonumber\\
&\leq \left( 1 - \mu \eta_k L \cdot \frac{ 1 - \beta^k }{ 1 - \beta } \right) \mathbb{E} \Big[ \big\Vert \Delta_{k} \big\Vert_\alpha^\alpha \Big]. 
\end{align}
Because $0 < \beta < 1$, when $k$ becomes large, we have $\beta^k \approx 0$ and hence we can approximately bound $Q_1$ as follows:
\begin{align} \label{equ:BndQ1_Mmnt_prim}
Q_1 \leq \left( 1 -  \frac{ \mu \eta_k L }{ 1 - \beta } \right) \mathbb{E} \Big[ \big\Vert \Delta_{k} \big\Vert_\alpha^\alpha \Big].
\end{align}

Secondly, we recursively use the Lemma~1 and bound $Q_2$ as follows:
\begin{align} \label{equ:BndQ2_Mmnt}
Q_2 &\leq \mathbb{E}\Big[ \big\Vert \sum_{ i=2 }^{k} \beta^{k-i} \boldsymbol{\xi}_i \big\Vert_\alpha^\alpha \Big] + 4 \beta^{ \alpha (k-1) } \cdot \mathbb{E}\Big[ \big\Vert \boldsymbol{\xi}_1 \big\Vert_\alpha^\alpha \Big]
\nonumber\\
&\leq \mathbb{E}\Big[ \big\Vert \sum_{ i=3 }^{k} \beta^{k-i} \boldsymbol{\xi}_i \big\Vert_\alpha^\alpha \Big] + 4 \beta^{ \alpha (k-2) } \cdot \mathbb{E}\Big[ \big\Vert \boldsymbol{\xi}_2 \big\Vert_\alpha^\alpha \Big] 
\nonumber\\
&\qquad \qquad \qquad \qquad \qquad \qquad \quad + 4 \beta^{ \alpha (k-1) } \cdot \mathbb{E}\Big[ \big\Vert \boldsymbol{\xi}_1 \big\Vert_\alpha^\alpha \Big]
\nonumber\\
&\leq \sum_{i=1}^k 4 \beta^{ \alpha (k-i) } \mathbb{E}\Big[ \big\Vert \boldsymbol{\xi}_i \big\Vert_\alpha^\alpha \Big] \leq \frac{ 4C }{ 1 - \beta^\alpha }.
\end{align}

Furthermore, we bound $Q_3$ as follows:
\begin{align} \label{equ:BndQ3_Mmnt}
&Q_3
\nonumber\\ 
&\stackrel{(a)}{\leq} d^{ 1 - \frac{1}{\alpha} } \cdot  \Big( \mathbb{E} \Big[ \big\Vert \sum_{ i=1 }^k \beta^{k-i} \sum_{n=1}^N ( h_{n, i} - \mu ) \nabla f_n ( \boldsymbol{w}_k ) \big\Vert_2^2 \Big] \Big)^{ \frac{\alpha}{2} } 
\nonumber\\
&= d^{ 1 - \frac{1}{\alpha} } \cdot \Big( \mathbb{E} \Big[ \sum_{n=1}^N \big\Vert \sum_{ i=1 }^k \beta^{k-i}  ( h_{n, i} - \mu ) \nabla f_n ( \boldsymbol{w}_k ) \big\Vert_2^2 \Big] \Big)^{ \frac{\alpha}{2} } 
\nonumber\\
& \leq d^{ 1 - \frac{1}{\alpha} } \cdot  \Big( N \mathbb{E} \Big[ \sum_{ i=1 }^k \beta^{2(k-i)}  ( h_{n, i} - \mu )^2 G^2 \Big] \Big)^{ \frac{\alpha}{2} } 
\nonumber\\
&= d^{ 1 - \frac{1}{\alpha} } N^{\alpha/2} \sigma^\alpha G^\alpha \left( \frac{ 1 - \beta^{2k} }{ 1 - \beta^2 } \right)^{ \frac{\alpha}{2} }
\nonumber\\
&\leq \frac{ d^{ 1 - \frac{1}{\alpha} } N^{\alpha/2} \sigma^\alpha G^\alpha }{ ( 1 - \beta^2 )^{ \frac{\alpha}{2} } }.
\end{align}
To this end, by replacing \eqref{equ:BndQ1_Mmnt_prim}, \eqref{equ:BndQ2_Mmnt}, and \eqref{equ:BndQ3_Mmnt} into \eqref{equ:Deltp1_Mmnt}, we have 
\begin{align}
\mathbb{E} \Big[ \big\Vert \Delta_{k+1} \big\Vert_\alpha^\alpha \Big] &\leq \left( 1 -  \frac{ \mu L  \eta_k }{ 1 - \beta } \right) \mathbb{E} \Big[ \big\Vert \Delta_{k} \big\Vert_\alpha^\alpha \Big] 
\nonumber\\
&\qquad \qquad  \quad + 4 \Big( \frac{ 4C }{ 1 - \beta^\alpha }  + \frac{ \sigma^\alpha G^\alpha d^{ 1 - \frac{1}{\alpha} } }{ N^{ \frac{\alpha}{2} }  (1-\beta^2)^{ \frac{\alpha}{2}  }  } \Big) \eta_k^\alpha
\end{align}
and the proof is completed by invoking Lemma~3 and removing the higher order terms as they become infinitesmall as $k$ goes large.

\subsection{Proof of Theorem~\ref{thm:GenErr}} \label{Apndx:GenErr_proof}
For ease of exposition, we denote $\hat{\mathcal{R}}(\boldsymbol{w}, \mathcal{D})$ by $\hat{\mathcal{R}}_{\mathcal{D}}( \boldsymbol{w} )$. For any $\boldsymbol{w} \in \mathbf{w}_{\mathcal{D}}$, we have the following
\begin{align} \label{equ:RD_bound}
&\vert \hat{\mathcal{R}}_{\mathcal{D}}( \boldsymbol{w} ) - \mathcal{R}(\boldsymbol{w}) \vert 
\nonumber\\
&=\! \vert \hat{\mathcal{R}}_{\mathcal{D}}( \boldsymbol{w} ) \!-\! \hat{\mathcal{R}}_{\mathcal{D}}( \boldsymbol{w}' ) \!+\! \hat{\mathcal{R}}_{\mathcal{D}}( \boldsymbol{w}' ) \! - \! \mathcal{R}(\boldsymbol{w}') \!+\! \mathcal{R}(\boldsymbol{w}') \! - \! \mathcal{R}(\boldsymbol{w}) \vert 
\nonumber\\
&\stackrel{(a)}{\leq} \! \vert \hat{\mathcal{R}}_{\mathcal{D}}( \boldsymbol{w}' )  - \mathcal{R}(\boldsymbol{w}') \vert + 2 \lambda \Vert \boldsymbol{w} - \boldsymbol{w}' \Vert 
\end{align}
where ($a$) follows from the Lipschitz property of the objective function. 

Because $\mathbf{w}_{ \mathcal{D} }$ is defined in $[0, 1]$ and initialized at $0$, it possesses a finite diameter almost surely. 
Therefore, we can consider a finite cover of $\mathbf{w}_{ \mathcal{D} }$ by balls of radii $s$ whereas the set of the centers of these balls is denoted as $J_s = J_s(\mathbf{w}_{ \mathcal{D} })$.
As such, $\forall \boldsymbol{w} \in \mathbf{w}_{ \mathcal{D} }$, $\exists \boldsymbol{w}' \in \mathbf{w}_{ \mathcal{D} }$ such that $\Vert \boldsymbol{w} - \boldsymbol{w}' \Vert \leq s$. 
By substituting this $\boldsymbol{w}'$ into \eqref{equ:RD_bound}, we have 
\begin{align}
\vert \hat{\mathcal{R}}_{\mathcal{D}}( \boldsymbol{w} ) - \mathcal{R}(\boldsymbol{w}) \vert \leq \vert \hat{\mathcal{R}}_{\mathcal{D}}( \boldsymbol{w}' )  - \mathcal{R}(\boldsymbol{w}') \vert + 2 \lambda s.
\end{align}
Furthermore, we take the supremum on both sides of the above inequality and arrive at the following
\begin{align} \label{equ:DifcntBnd}
\sup_{ \boldsymbol{w} \in \mathbf{w}_{ \mathcal{D} } } \!\! \vert \hat{\mathcal{R}}_{\mathcal{D}}( \boldsymbol{w} ) - \mathcal{R}(\boldsymbol{w}) \vert \leq \max_{ \boldsymbol{w} \in J_s } \vert \hat{\mathcal{R}}_{\mathcal{D}}( \boldsymbol{w} ) - \mathcal{R}(\boldsymbol{w}) \vert + 2 \lambda s.
\end{align}
Given any $\varepsilon > 0$, we have 
\begin{align}
&\mathbb{P} \left( \max_{ \boldsymbol{w} \in J_s } \vert \hat{\mathcal{R}}_{\mathcal{D}}( \boldsymbol{w} ) - \mathcal{R}(\boldsymbol{w}) \vert \geq \varepsilon \right) 
\nonumber\\
&= \mathbb{P} \left( \cup_{ \boldsymbol{w} \in J_s } \left\{  \vert \hat{\mathcal{R}}_{\mathcal{D}}( \boldsymbol{w} ) - \mathcal{R}(\boldsymbol{w}) \vert \geq \varepsilon \right\} \right)
\nonumber\\
&\leq \sum_{ \boldsymbol{w} \in J_s } \mathbb{P} \left( \vert \hat{\mathcal{R}}_{\mathcal{D}}( \boldsymbol{w} ) - \mathcal{R}(\boldsymbol{w}) \vert \geq \varepsilon \right)
\nonumber\\
&\stackrel{(a)}{\leq} 2 \vert J_s \vert \cdot \exp\left( - \frac{ 2 \vert \mathcal{D} \vert \varepsilon^2 }{ B^2 } \right)
\end{align}
where ($a$) follows from the Hoeffding's inequality \cite{Hoe:63}. We assign $p = 2 \vert J_s \vert \exp( - 2 \vert \mathcal{D} \vert \varepsilon^2 / B^2 )$. Then, with probability at least $1-p$, the following holds
\begin{align}
\max_{ \boldsymbol{w} \in J_s } \vert \hat{\mathcal{R}}_{\mathcal{D}}( \boldsymbol{w} ) - \mathcal{R}(\boldsymbol{w}) \vert \leq \varepsilon.
\end{align}
By rewriting $\varepsilon$ in terms of $p$ and leveraging \eqref{equ:DifcntBnd}, we have 
\begin{align}
\sup_{ \boldsymbol{w} \in \mathbf{w}_{ \mathcal{D} } } \!\! \vert \hat{\mathcal{R}}_{\mathcal{D}}( \boldsymbol{w} ) - \mathcal{R}(\boldsymbol{w}) \vert \leq B \sqrt{ \frac{ \log(2 J_s) + \log(1/p) }{ 2 \vert \mathcal{D} \vert } } + 2 \lambda s.
\end{align}
Furthermore, given an arbitrary sequence $\{ s_k \}_{ k \in \mathbb{N} }$ whereas $\lim_{ k \rightarrow \infty } s_k = 0$, for $\varepsilon > 0$, there exists $k_\varepsilon > 0$ such that for $k \geq k_\varepsilon$ the following holds 
\begin{align} \label{equ:JsBnd}
\log(J_{s_k}) \leq \left( \overline{ \mathrm{dim}_{\mathrm{M}} } \mathbf{w}_{ \mathcal{D} } + \varepsilon \right) \cdot \log( s_k^{-1} )
\end{align}
in which $\overline{ \mathrm{dim}_{\mathrm{M}} } E$ represents the upper Minkovsky dimension of a Borel set $E$.
To this end, by assigning $\varepsilon = \overline{ \mathrm{dim}_{\mathrm{M}} } \mathbf{w}_{ \mathcal{D} }$ in \eqref{equ:JsBnd} and having $s_k = 1/\sqrt{ \lambda^2 \vert \mathcal{D} \vert }$, we have 
\begin{align}
&\sup_{ \boldsymbol{w} \in \mathbf{w}_{ \mathcal{D} } } \!\! \vert \hat{\mathcal{R}}_{\mathcal{D}}( \boldsymbol{w} ) - \mathcal{R}(\boldsymbol{w}) \vert 
\nonumber\\
&\leq B \sqrt{ \frac{ \log(2) + \overline{ \mathrm{dim}_{\mathrm{M}} } \mathbf{w}_{ \mathcal{D} } \cdot \log( \lambda^2 \vert \mathcal{D} \vert ) + \log(1/p) }{ 2 \vert \mathcal{D} \vert } } + \frac{ 2 }{ \vert \mathcal{D} \vert }
\nonumber\\
&\stackrel{(a)}{\leq} B \sqrt{ \frac{ 2 \, \overline{ \mathrm{dim}_{\mathrm{M}} } \mathbf{w}_{ \mathcal{D} } \cdot \log( \lambda^2 \vert \mathcal{D} \vert ) + \log(1/p) }{ \vert \mathcal{D} \vert } }
\end{align}
where ($a$) holds when $\vert \mathcal{D} \vert$ is sufficiently large. 
Finally, the proof is completed by noticing the fact that $\overline{ \mathrm{dim}_{\mathrm{M}} } \mathbf{w}_{ \mathcal{D} } = { \mathrm{dim}_{\mathrm{H}} } \mathbf{w}_{ \mathcal{D} } \leq \alpha$.

\end{appendix}
\bibliographystyle{IEEEtran}
\bibliography{bib/StringDefinitions,bib/IEEEabrv,bib/howard_SGD_HeavTail}

\end{document}

%% file: SupportDocuments/aconym.tex
\acrodef{CCDF}{complementary cumulative distribution function}
\acrodef{CF}{characteristic function}
\acrodef{PPP}{Poisson point processe}
\acrodef{RV}{random variable}
\acrodef{i.i.d.}{independent and identically distributed}
\acrodef{PDF}{probability distribution function}
\acrodef{CDF}{cumulative distribution function}
\acrodef{ch.f.}{characteristic function}
\acrodef{AWGN}{additive white Gaussian noise}
\acrodef{SNR}{signal-to-noise ratio}
\acrodef{LRT}{likelihood ratio test}
\acrodef{DRT}{distance ratio test}
\acrodef{GLRT}{generalized likelihood ratio test}
\acrodef{CRLB}{Cram\'{e}r-Rao lower bound}
\acrodef{CRB}{Cram\'{e}r-Rao bound}
\acrodef{ZZLB}{Ziv-Zakai lower bound}
\acrodef{ZZB}{Ziv-Zakai bound}
\acrodef{LOS}{line-of-sight}
\acrodef{ToF}{time-of-flight}
\acrodef{NLOS}{non-line-of-sight}
\acrodef{GDOP}{geometric dilution of precision}
\acrodef{GPS}{Global Positioning System}
\acrodef{FIM}{Fisher information matrix}
\acrodef{PEB}{position error bound}
\acrodef{SPEB}{squared position error bound}
\acrodef{TOA}{time-of-arrival}
\acrodef{TOF}{time-of-flight}
\acrodef{WSN}{wireless sensor network}
\acrodef{MAC}{medium access control}
\acrodef{RSS}{received signal strength}
\acrodef{WAF}{wall attenuation factor}
\acrodef{TDOA}{time difference-of-arrival}
\acrodef{RF}{radiofrequency}
\acrodef{RTT}{round-trip time}
\acrodef{AOA}{angle-of-arrival}
\acrodef{MF}{matched filter}
\acrodef{ED}{energy detector}
\acrodef{ML}{maximum likelihood}
\acrodef{MSE}{mean-square error}
\acrodef{RMSE}{root-mean-square error}
\acrodef{LEO}{localization error outage}
\acrodef{ppm}{part-per-million}
\acrodef{ACK}{acknowledge}
\acrodef{UWB}{Ultrawide bandwidth}
\acrodef{TNR}{threshold-to-noise ratio}
\acrodef{LS}{least squares}
\acrodef{IR-UWB}{impulse radio UWB}
\acrodef{FCC}{Federal Communications Commission}
\acrodef{TH}{time-hopping}
\acrodef{PPM}{pulse position modulation}
\acrodef{MUI}{multi-user interference}
\acrodef{PDP}{power delay profile}
\acrodef{BPZF}{band-pass zonal filter}
\acrodef{SIR}{signal-to-interference ratio}
\acrodef{SINR}{signal-to-interference-plus-noise ratio}
\acrodef{RFID}{radio frequency identification}
\acrodef{WPAN}{wireless personal area network}
\acrodef{WWB}{Weiss-Weinstein bound}
\acrodef{DP}{direct path}
\acrodef{MF}{matched filter}
\acrodef{MMSE}{minimum-mean-square-error}
\acrodef{SBS}{serial backward search}
\acrodef{SBSMC}{serial backward search for multiple clusters}
\acrodef{NBI}{narrowband interference}
\acrodef{WBI}{wideband interference}
\acrodef{INR}{interference-to-noise ratio}
\acrodef{CR}{channel response}
\acrodef{CIR}{channel impulse response}
\acrodef{CR}{channel  response}
\acrodef{RADAR}{radar}
\acrodef{MUR}{Multistatic radar}
\acrodef{JBSF}{jump back and search forward}
\acrodef{HDSA}{high-definition situation-aware}
\acrodef{RRC}{root raised cosine}
\acrodef{ST}{simple thresholding}
\acrodef{BTB}{Bellini-Tartara bound}
\acrodef{P-Max}{$P$-Max}  
\acrodef{MIMO}{multiple-input multiple-output}
\acrodef{MAP}{maximum a posteriori}
\acrodef{FG}{factor graph}
\acrodef{OP}{outage probability}
\acrodef{WED}{wall extra delay}
\acrodef{RMS}{root mean square}
\acrodef{SPAWN}{sum-product algorithm over a wireless network}
\acrodef{MDD}{minimum distance distribution}
\acrodef{MAP}{maximum a posteriori probability}
\acrodef{SAP}{small cell access point}
\acrodef{UE}{user equipment}
\acrodef{MBS}{macro cell base station}
\acrodef{UER}{\ac{UE} Relay}
\acrodef{D2D}{device-to-device}
\acrodef{MBS}{macro base station}
\acrodef{CSI}{channel state information}
\acrodef{OGR}{outage guard region}
\acrodef{FUR}{feasible UER region}
\acrodef{EHR}{energy harvesting region}
\acrodef{EH}{energy harvesting}
\acrodef{D2D-EHSN}{D2D communication provided \ac{EH} small cell network}
\acrodef{D2D-EHHN}{D2D communication provided \ac{EH} heterogeneous network}
\acrodef{3GPP}{3rd Generation Partnership Project}
\acrodef{BS}{base station}
\acrodef{DF}{decode and forward}
\acrodef{CCDF}{complementary cumulative distribution function}
\acrodef{ZF}{zero forcing}
\acrodef{RZF}{regularized zero forcing}
\acrodef{WLLN}{weak law of large number}
\acrodef{SLLN}{strong law of large numbers}
\acrodef{TDD}{Time-division duplex}
\acrodef{EE}{energy efficiency} 
\acrodef{HetNet}{heterogeneous network} 
\acrodef{SCP}{Single Cell Processing}
\acrodef{CBF}{Coordinated Beamforming}

%% file: SupportDocuments/defmetric.tex
\usepackage{color}
\usepackage{dsfont}
\usepackage{bbm}

\DeclareMathAlphabet{\mathsf}{OML}{cmbr}{m}{it}

\newtheorem{definition}{\bf Definition}
\newtheorem{theorem}{\bf Theorem}
\newtheorem{lemma}{\bf Lemma}
\newtheorem{corollary}{\bf Corollary}

\newtheorem{remark}{\bf Remark}

\newtheorem{assumption}{\bf Assumption}

\newcommand{\bd}{\begin{description}}
\newcommand{\ed}{\end{description}}
\newcommand{\be}{\begin{enumerate}}
\newcommand{\ee}{\end{enumerate}}
\newcommand{\bi}{\begin{itemize}}
\newcommand{\ei}{\end{itemize}}
\newcommand{\bl}{\begin{list}}
\newcommand{\el}{\end{list}}
\newcommand{\bt}{\begin{tabbing}}
\newcommand{\et}{\end{tabbing}}